\documentclass[twocolumn]{aastex62}
\usepackage{textcomp}
\usepackage{natbib}
\usepackage{mathtools}
\usepackage{hyperref}
\usepackage{amssymb}
\usepackage{graphicx}
\usepackage{lineno}
\newcommand{\quotes}[1]{``#1''}

\graphicspath{{./}{figures/}}

\begin{document}
\title{Point source contribution to the Diffuse X-ray Background below 1 keV and its effect on our understanding of the circum-galactic medium}

\author{Sicong Huang}
\affiliation{University of Miami, Department of Physics, Coral Gables, FL 33124, USA}

\author[0000-0002-1697-186X]{Nico Cappelluti}
\affiliation{University of Miami, Department of Physics, Coral Gables, FL 33124, USA}

\author[0000-0003-3702-7592]{Massimiliano Galeazzi}
\affiliation{University of Miami, Department of Physics, Coral Gables, FL 33124, USA}

\author[0000-0003-1880-1474]{Anjali Gupta}
\affiliation{Biological and Physical Sciences, Columbus State Community College, Columbus, OH 43215}

\author[0000-0002-0924-9668]{Wenhao Liu}
\affiliation{Purple Mountain Observatory, Chinese Academy of Science, Nanjing 210034, People's Republic of China}

\author[0000-0002-2567-2036]{Eugenio Ursino}
\affiliation{Purdue University Fort Wayne, Department of Physics, Fort Wayne, IN 46805}

\author{Tomykkutty J. Velliyedathu}
\affiliation{University of Miami, Department of Physics, Coral Gables, FL 33124, USA}

\correspondingauthor{Massimiliano Galeazzi}
\email{galeazzi@miami.edu}

%%%%%%%%%%%%%%%%%%%%%%%%%%%%%%%%%%%%%%%%%%%%--ABSTRACT--%%%%%%%%%%%%%%%%%%%%%%%%%%%%%%%%%%%%%%%%%%%%%%%%%%%%%%%%%%
\begin{abstract}\label{abstract}

We studied the spectral signature of different components of the Diffuse X-ray Background (DXB), including Local Hot Bubble (LHB), Solar Wind Charge Exchange (SWCX), Galactic Halo, and typically unresolved point sources (galaxies and AGN), in the direction of the Chandra Deep Field South (CDFS) using the 4 Ms XMM-Newton survey and Chandra 4 Ms Source Catalog. In this paper, we present our results showing how the different components contribute to the DXB below 1 keV. 
In particular, we have found that $\sim6\%$ of the emission at \(\frac{3}{4}\) keV (all-sky average value $\approx 3\times10^{-3}$ cm$^{-6}$pc), which is typically associated with Galactic Halo (GH) and Circum-galactic medium (CGM) is, in fact, due to emission from typically unresolved galaxies.
 We will discuss the effect that this has on our understanding of GH and CGM, and to our understanding of the missing CGM baryons.

\end{abstract}

%%%%%%%%%%%%%%%%%%%%%%%%%%%%%%%%%%%%%%%%%%%%--INTRODUCTION--%%%%%%%%%%%%%%%%%%%%%%%%%%%%%%%%%%%%%%%%%%%%%%%%%%%%%%%%%%
\section{Introduction}\label{introduction}
The presence of a diffuse X-ray glow in the sky, what we now call the Diffuse X-ray Background (DXB), was already evident in the very first extra-solar X-ray detection by \cite{Giacconi1962}. The origin and nature of such glow have been studied since and it is now believed that, above 1 keV, the DXB is isotropic and 80\% of the emission can be resolved into discrete point sources, mainly Active Galactic Nucleus (AGN), plus galaxies and stars. Below 1 keV, the nature of the DXB becomes more complicated with contributions from several overlapping components, including \quotes{local}, galactic and extragalactic sources \citep{Snowden1998}. The local emission originates within roughly 100 pc from the Sun and consists mostly of the emission from Local Hot Bubble (LHB), with a significant contribution from Solar Wind Charge Exchange (SWCX) \citep{Galeazzi2014Nature}. The properties and structure of the LHB have been recently studied and constrained by \cite{Liu2017}, \cite{Wulf2019} and \cite{Pelgrims2020}, providing us with a chance to quantitatively study the remaining DXB emission.

Although significant efforts have already been dedicated to the study of SWCX \citep{Uprety2016, Kuntz2019, Wulf2019}, it is still a challenge to quantify its contribution to the DXB. SWCX occurs when highly charged ions in the Solar Wind charge-exchange with neutral atoms in the solar system, both in the heliosphere (heliospheric emission) and in Earth's exosphere (geocoronal emission). SWCX emission is a major contribution to the foreground X-ray flux and often considered problematic because its temporal variation makes it difficult to model, and its similarity with LHB and Galactic Halo emission in spectral features makes it hard to disentangle it from the astrophysical components. Thus, modeling SWCX contamination is a challenging but necessary task in this work in order to fully disentangle contribution from different components. Although it is arduous to determine the amount of SWCX emission due to its wide variation with time and view-angle, as well as uncertainties in relative abundances, densities, and velocities in ions, it is possible to obtain reasonable constraints on the emission lines of the two most important ions for XMM and Chandra -- O VII and O VIII -- and deduce the extent of the SWCX contamination in the direction of the investigation. In this paper we use two different approaches for SWCX to model and investigate its temporal variation with the solar cycle phase and its relative contribution to DXB.  

Besides LHB and SWCX, an additional thermal emission, being absorbed by galactic hydrogen, is traditionally thought to originate from the Galactic Halo (GH) or Circum-Galactic Medium (CGM) \citep{Snowden1998, Kuntz+Snowden2000, Smith2007, Galeazzi2007, Gupta2012, Gupta2014, Gupta2017}. 

Extra-galactic emission includes emission from unresolved point sources and the Warm-Hot Intergalactic Medium (WHIM) \citep{Mushotzky2000, Croft2001, Ursino+Galeazzi2006, Galeazzi2009, Nico2016} 

The population of unresolved point sources is typically dominated by Active Galactic Nuclei (AGN) and their cumulative spectrum is well approximated, above 1 keV, by an absorbed powerlaw. However, this simple model is less reliable for other types of point sources, with significant effects below  keV.
For observations toward the galactic plane, stars (with an additional thermal component) play a significant role \cite{Hands2004}. However, in off-plane observation, like the one we are considering, their contribution is negligible. Galaxies are also a non-negligible component and, like stars, their spectra include a significant thermal component, so that some of the absorbed thermal emissions typically associated with GH/CGM may be extragalactic in origin instead. This is particularly important because, although the GH/CGM has been extensively studied observationally, its nature is still largely uncertain, which leads to the demand for constraints on the theoretical models of GH/CGM. For example, the extent and the mass of the hot gas in GH/CGM can be constrained by combining absorption and emission spectral features \citep{Gupta2012}. 
Knowing quantitatively how much the unresolved galaxies contribute to the absorbed thermal emission can help us further constrain the structure of the GH/CGM, and how much the hot gas contributes to our Galaxy's baryons.

The Chandra Deep Field South (CDFS) is one of the deepest X-ray field covered by both Chandra and XMM-Newton. Given the high flux limits achieved by Chandra and XMM-Newton, most of the DXB and point sources, especially AGN, are virtually completely resolved \citep{Gilli2007}. 
\cite{Hickox2006} concluded that the contribution of resolved sources in CDFS accounts for
$\sim 80$\% of the 0.5-8 keV Chandra flux. The CDFS has also been covered extensively at other wavelenghts, giving us a deep understanding of the nature of those sources. 
As AGN dominates in flux above 1 keV, most of the work on the CDFS is dedicated to obscured and unobscured AGN in the hard band \citep{Vito2013, Comastri2014, Corral2019, Iwasawa2020}. Nevertheless, it is worth noting that, below 1 keV, the uncertainty of measuring the DXB can be significantly reduced with the availability of large source samples along with long exposure times ($\sim$ 4Ms) in CDFS, which provides a great opportunity to obtain better constraints on the DXB parameters.
\cite{Hickox2006} measured the unresolved 0.5-8 keV DXB in Chandra Deep Field South and Chandra Deep Field North (CDFN) using the Chandra 1 and 2 Ms exposures, respectively. In their work, they used Chandra data to generate spectra and fitted with 
the APEC thermal plasma model\citep{Smith2001} for soft diffuse emission and an absorbed powerlaw for point sources emission. In this work, we focused on CDFS and followed a similar procedure with improvements to produce and fit the spectra for both DXB and galaxies. Most importantly, we used data from XMM-Newton to take advantage of its better performance below 1 keV, including effective area, spectral resolution, signal-to-noise ratio, and accuracy in estimation of instrumental background. 
We used a composite model with parameters for LHB set to ROSAT observations to fit the spectra and studied the contributions from different components in the soft diffuse emission. To fully remove the contribution from point sources, we utilize the source catalog from Chandra, acknowledging its high angular resolution in detecting point sources with greater sensitivity and minimal source confusion. 

In analyzing the contribution from point sources, the high statistics available from CDFS and the availability of data from Chandra and other multi-wavelength observatories allows us to go to a much lower flux threshold than for any other mission/target. In particular, in this work, the focus is on the contribution to the DXB from unresolved AGN. In this regard, we focused on how the contribution depends on the flux limit of a pointing by dividing the AGN into multiple flux bands and analyzing their spectral properties separately. 

This paper is structured as follows: Section 2 describes the stage of data reduction, section 3 describes the stage of data analysis, and in section 4 we show the results with discussion and conclusions.
%%%%%%%%%%%%%%%%%%%%%%%%%%%%%%%%%%%%%%%%%%%%--DATA Preparation--%%%%%%%%%%%%%%%%%%%%%%%%%%%%%%%%%%%%%%%%%%%%%%%%%%%%%%%%%%
\section{Data Reduction}\label{Data Preparation}
To clean the data and generate spectra, we used the XMM-Newton Extended Source Analysis Software package (XMM-ESAS, sasversion: 19.1.0) as the main data analysis tool and followed the standard data process procedure for XMM-Newton EPIC observations of extended objects and the diffuse background \citep{XMM Cookbook}, with modifications such as manually modifying the event lists and generating external source lists. Other tools used in this work include XMM-Newton Science Analysis System (SAS, sasversion: 19.1.2, with corresponding calibration data), the X-ray Spectral Fitting Package (XSPEC, version 11) with the atomic data for astrophysicists database (AtomDB, version 3.0.9), and FTOOLS \footnote{\url{http://heasarc.gsfc.nasa.gov/ftools}} \citep{Blackburn1999}.
%%%%%%%%%%%%%%%%%%%%%%%%%%%%%%%%%%%%%%%%%%%%--Field Selection--%%%%%%%%%%%%%%%%%%%%%%%%%%%%%%%%%%%%%%%%%%%%%%%%%%%%%%%%%%

\subsection{Field Selection}\label{Field Selection}
For the purpose of this work, we chose the Chandra Deep Field South field, which is one of the deepest multi-wavelength observation fields covering 440 square arcminutes. It is centered at RA = 03h32m28.2s, DEC = -27d48m36.0s, with a relatively low Galactic neutral hydrogen column density (\(\textrm{N}_\textrm{H} \simeq 0.0088 \times 10^{22}\) cm\(^{-2}\), \cite{Dickey1990}) and no bright features. 

%\ref{tab:osbservations} 

To study the spectral features of both DXB and galaxies, we make use of 33 XMM-Newton observations of the CDFS with a nominal exposure time of 3.45 Ms. Table 1 lists the exposure time and center coordinate for each observation. Notice that only 32 observations are listed and one observation is left out due to low exposure. For each observation, the aim point is slightly different due to the shifts in the satellite position. By manually defining the extraction region (see Section \ref{Spectra generation}), we made sure that all spectra extracted for 32 observations are in the same region with the same center. 
Among the three imaging detectors aboard XMM, the EPIC PN detector is more sensitive and has a larger effective area than the two EPIC MOS detectors. In this work, we chose to use the data only from the PN camera. 
For this work, we also limited our analysis to the portion of the XMM-Newton field that is also covered by Chandra, to have a uniform source removal across the field.

%%%%%%%%%%%%%%%%%%%%%%%%%%%%%%%%%%%%%%%%%%%%--Source Exclusion--%%%%%%%%%%%%%%%%%%%%%%%%%%%%%%%%%%%%%%%%%%%%%%%%%%%%%%%%%%

\subsection{Point Sources Removal}\label{Source Exclusion}
The \verb|ESAS| script \verb|cheese| 
generates a list of detected sources
and a cheese mask image with all the XMM-detected sources removed  (Figure \ref{fig:dxb_cheese}). However, since the Chandra observatory has a better angular resolution and point spread function (PSF) than XMM-Newton, it has the ability to form sharper images and detect fainter sources. To fully separate the emission of DXB and point sources, the Chandra Deep Field South 4-Megasecond (CDFS4Ms) Catalog \citep{Chandra4Ms} is used. We note that, while we looked at the deeper Chandra 7-Megasecond Catalog \citep{Chandra7Ms}, when coupled with XMM-Newton PSF, the larger number of sources could not be used due to source overlap. As our analysis has shown that the contribution to the diffuse flux from residual sources is negligible, we decided to use the smaller catalog. The catalog contains 740 sources, along with their coordinates, redshifts, full-band fluxes, and classifications. We reformatted the catalog and added it to the source lists generated by \verb|cheese|, along with 46 extended sources (galaxy clusters) detected by \cite{FinoguenovClusters}. The radii of XMM-detected sources are automatically calculated by \verb|cheese|. To determine the cut-off radii of point sources from CDFS4MS catalog, we used the XMM-Newton half energy width (HEW), propagated to off axis sources using a linear relationship to determine the radius r (in arcseconds) for source extraction (based on \quotes{XMM-Newton Users Handbook}, Issue 2.17, 2019 (ESA: XMM-Newton SOC)):
\begin{equation} \label{eq:r50}
r = \frac{1}{2} \times \textup{HEW} + k \times \theta
\end{equation}
where $k = 1.27 arcsec\times arcmin^{-1}$ 
%derived from Figure 2 
and $\theta$ is the off-axis angle in arcminutes.

For sources classified as galaxies, we decided to use both the HEW (same as Equation \ref{eq:r50}) and a larger radius with 90\% encircled energy for different spectra sets:
\begin{equation} \label{eq:r90}
r = \textup{W90} + k \times \theta\
\end{equation}
where W90 is the on-axis 90\% encircled energy radius. 

In this way, two sets of DXB spectra can be generated: one set removes 50\% energy of AGN and 50\% energy of galaxies (DXB-50 hereafter), and another set removes 50\% energy of AGN and 90\% energy of galaxies (DXB-90 hereafter). We note that removing 90\% for all sources is impossible, as it would remove essentially all data. As we'll discuss later, we decided to apply the larger radius to galaxies due to their additional thermal component, which is not present in AGN. 

\subsection{Soft Proton (SP) Flaring}\label{soft proton (SP) flaring}

Low energy protons in the Van Allen belt can enter the telescope and interact with the Wolter-I optics aboard XMM-Newton in forms of sudden flares in count rate in the EPIC instrument,  causing lost of observing time from hundreds of seconds to hours \citep{Fioretti2016}. A filtering algorithm for SP flaring must be run to remove the contamination. The \verb|ESAS| routines \verb|epchain| and \verb|pn-filter| filter the event files from SP flaring contamination by creating light-curves and count rate histograms, and based on light curve, it generates cleaned event files using only time intervals where the FOV count rate is below 1.2 times the count rate of the unexposed corners. We carefully inspected the count rate histograms and confirmed that after filtering, the effect of residual SP contamination becomes minimal for our analysis. 

\subsection{Spectra Production}\label{Spectra generation}
The wealth of information available from the CDFS field allows us to generate individual spectra for AGNs, Galaxies, and the diffuse emisssion. However, the XMM-Newton PSF limits the number of sources that can be considered, as a number of sources too high will create source overlap in the point spectra and would leave no ``empty'' regions for the extraction of the diffuse emission. After some testing, a good compromise that gave us sufficient solid angle for the analysis of the diffuse emission, while extracting the required information regarding point sources has been the use of the Chandra 4-Megasecond catalog and a circular region around point sources corresponding to XMM-Newton Half Power Diameter (50\% of the flux). We separately tested both the use of the large Chandra 7-Megasecond catalogue and larger areas around point sources, but both approaches lead to source mixing and not enough solid angle for the diffuse analysis.
As a check that the most relevant point sources have been identified in the investigation, we confirmed that the power law component in the diffuse emission is consistent with the one that we obtained looking at all sources, which is what we expected considering that we used the half power diameter to remove point sources from diffuse emission and to generate the point source spectra. 

During the analysis of spectra from galaxies, we noticed the presence of an additional component, consistent with thermal emission, that could be attributed both to the galaxies themselves, or contamination from the diffuse emission in the galaxy extraction regions. To properly identify those components, we therefore decided to increase the extraction radius around galaxy to 90\% of the power. We chose 90\% of the power as we wanted to include (or remove, depending on the spectra) as much contribution from galaxies as possible, without making the regions too large. This was possible as the number of galaxies is significantly smaller than that of AGNs.  A total number of four spectral sets are therefore generated. For DXB, two sets of spectra are generated: one with 50$\%$ of AGN emission plus 50$\%$ of galaxy emission removed, and the other with 50$\%$ of AGN emission plus 90$\%$ of galaxy emission removed. Similarly, two sets of spectra of galaxies are generated: one with 50$\%$ of galaxy emission selected (Galaxy-50 hereafter) and the other with 90$\%$ selected (Galaxy-90 hereafter). Since the diffuse emission scales differently than the point source one for the two sets of radii, by calculating and comparing the normalization of the additional component from all four sets, we are able to determine its origin.

\subsubsection{DXB Spectra}
\begin{figure}
\begin{center}
\includegraphics[width=6cm]{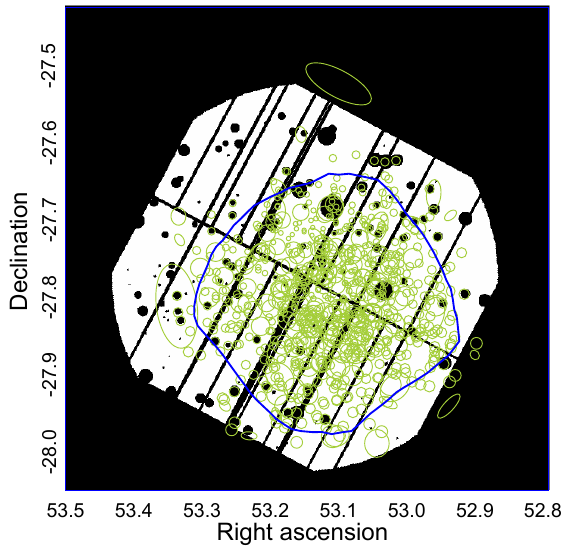}
\end{center}
\vspace{-0.2cm}
\caption{Cheese map for DXB-90 for ObsID 0555780101. The black regions inside XMM FOV are point sources detected by XMM cheese and CCD gaps; the green circles are point sources from the Chandra source catalog; the green ellipses are extended sources from \cite{FinoguenovClusters}; and the blue polygon is the data selection region that covers the Chandra FOV.}
\label{fig:dxb_cheese}
%\vspace{-12pt}
\end{figure}

We generated the DXB spectra for 32 XMM pointings using the \verb|ESAS| script \verb|pn-spectra| with the \verb|cheese| masks to remove all the contribution from point sources (Figure \ref{fig:dxb_cheese}). The corresponding Response Matrix Files (RMFs) and  Ancillary Response Files (ARFs) are also generated in this process. As discussed before, we limited our spectra extraction region to a polygon that covers Chandra FOV, and contains most of the Chandra-detected sources. The area was chosen by selecting the region where the Chandra total exposure is at least 30\% percent of the maximum exposure, to guarantee relatively uniform source detection. Since the Chandra field is smaller than the XMM-Newton one, this also guarantees relative uniform XMM-Newton exposure.

A crucial part of the data reduction is the accurate modeling of the instrumental background, or \quotes{quiescent particle background} (QPB) 
\citep{Kuntz+Snowden2008}. This internal background is typically induced by interaction between high energy cosmic particles and the satellite material, showing both spatial and temporal variations.
We used \verb|XMM-ESAS| to model the QPB for both spectral and spatial analysis of EPIC pn observations using data observed by unexposed pixels in the corners.
This is achieved by \verb|ESAS| script \verb|pn_back|. 
The script uses a combination of Filter Wheel Closed (FWC) data and data from the corners of CCDs to capture the spatial and temporal variation of the QPB. 
Between 1.3 keV and 1.75 keV, the strong Al K$\alpha$  fluorescent instrumental line ($\sim$ 1.49 keV) can affect the data significantly. This part cannot be modeled properly using FWC data and 
we decided to exclude this region. 

Since the XMM-Newton effective area in the 10-14 keV band is negligible, the count rate in this band is mostly due to QPB. We scaled the normalization of QBP such that the count rate in the QPB-subtracted spectrum is approximately 0 in 10-14 keV range to get an accurate model of QPB \citep{Hickox2006}. 
Finally, using \verb|FTOOLS| task \verb|grppha|, the DXB spectra are binned with a minimum of 25 counts per channel. In this process, the source spectra are associated with corresponding ARFs, RMFs and instrumental background files.
%%%%%%%%%%%%%%%%%%%%%%%%%%%%%%%%%%%%%%%%%%%%--Figure 4--%%%%%%%%%%%%%%%%%%%%%%%%%%%%%%%%%%%%%%%%%%%%%%%%%%%%%%%%%%
%%%%%%%%%%%%%%%%%%%%%%%%%%%%%%%%%%%%%%%%%%%%--Figure 4 END--%%%%%%%%%%%%%%%%%%%%%%%%%%%%%%%%%%%%%%%%%%%%%%%%%%%%%%%%%%

\subsubsection{Galaxy Spectra}
\begin{figure}
\begin{center}
\includegraphics[width=6cm]{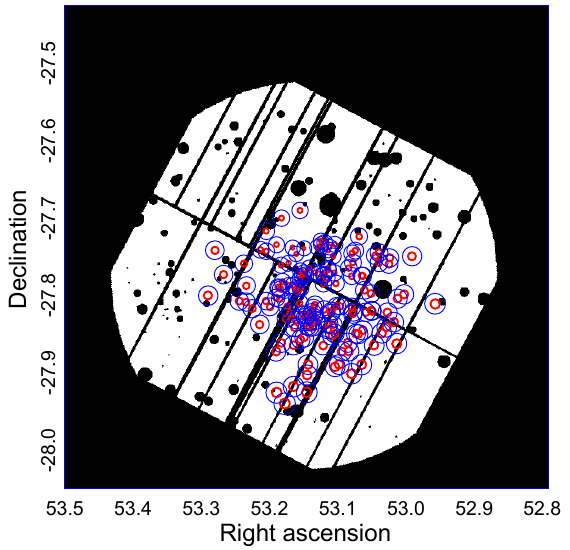}
\end{center}
\vspace{-0.2cm}
\caption{Cheese map for ObsID 0555780101. The red circles are Chandra-only detected galaxies with 50\% power diameter. The blue circles are galaxies with 90\% power diameter. As the figure shows, the overlapping areas for Galaxy-90 is much larger than Galaxy-50.}
\label{fig:gal_cheese}
\end{figure}

We selected sources in Chandra4MS catalog labeled as \quotes{GALAXY} to produce the spectra for galaxies. We limited the spectra extraction regions for galaxies inside the same polygons as DXB to avoid large off-axis angles. Since our goal is to compare the spectra of DXB with unresolved galaxies, namely the galaxies not detected by XMM-Newton, we excluded galaxies whose centers lie inside XMM-detected sources. We also excluded sources inside galaxies clusters, as those sources tend to be very bright and may introduce large uncertainties to our results.  
Eventually, we selected around 140 sources (Figure \ref{fig:gal_cheese}). To generate spectra for galaxies, we used two different treatments for galaxies with 50\% energy diameter (Galaxy-50) and 90\% energy diameter (Galaxy-90). For Galaxy-50, we used \verb|XMM-SAS| script \verb|evselect| to select a circular region around each galaxy, using the radius defined in Equation \ref{eq:r50}. A spectrum is then extracted in each region by setting the parameter \verb|spectrumset| to \quotes{True}. For each observation, we used \verb|FTOOLS| task \verb|mathpha| to combine the spectra from individual galaxies into a single stacked spectrum. We used \verb|SAS| script \verb|arfgen| and \verb|rmfgen| to create ARF and RMF for each galaxy. Then we used \verb|addarf| and \verb|addrmf| to combine ARFs and RMFs into a single ARF file and RMF file to be grouped with the stacked spectrum for each observation. This step is valid given that the overlapping area of neighboring galaxies is insignificant compared to the total area. We visually inspected the galaxy selection region for each observation and made sure this is the case. We used the same instrumental background files for DXB to group with the stacked spectrum, ARF and RMF using \verb|grppha|.

When dealing with Galaxy-90, it is not uncommon that galaxies overlap with each other. In order to avoid over-counting the number of photons in the overlapping regions, it is required to use a different strategy to generate this spectrum set. Because the overall solid angle of selected galaxies is somewhat comparable with DXB, the whole region that contains all galaxies can be treated as a diffuse emission. We decided to use the \verb|ESAS| script \verb|pn-spectra| for spectrum generation, similarly to what we did for DXB. The \verb|pn-spectra| routine utilizes cleaned event files produced by \verb|epchain| and selects interested events inside a user-defined region. Because the total galaxy region is combined from $\sim 140$ circular regions and in a highly irregular shape, \verb|pn-spectra| is unable to handle such a complicated region. In this case, we decided to modify the event files directly by manually selecting events inside the galaxy region and removing the others. The selected events produce the same result as using a region file. Then we used \verb|pn-spectra| with this modified event files and generated a spectrum for each observation, along with the corresponding ARFs and RMFs.

%%%%%%%%%%%%%%%%%%%%%%%%%%%%%%%%%%%%%%%%%%%%--Figure 5--%%%%%%%%%%%%%%%%%%%%%%%%%%%%%%%%%%%%%%%%%%%%%%%%%%%%%%%%%%
%%%%%%%%%%%%%%%%%%%%%%%%%%%%%%%%%%%%%%%%%%%%--Figure 5 END--%%%%%%%%%%%%%%%%%%%%%%%%%%%%%%%%%%%%%%%%%%%%%%%%%%%%%%%%%%

\subsubsection{AGN Spectra}\label{AGN Spectra creation}
The AGN extraction follows a similar procedure as Galaxy-50 for spectra production. We used \verb|XMM-SAS| script \verb|evselect| to select AGN in CDFS4MS catalog with radii calculated from Equation \ref{eq:r50} and generated spectra using the \verb|spectrumset| parameter. For each observation, the AGN spectra are divided and put into 6 different categories based on their soft-band fluxes (see Table 5). The spectra in each category are then merged by \verb|mathpha|. The corresponding RMFs and ARFs are generated using \verb|rmfgen| and \verb|arfgen|, then merged by \verb|addrmf| and \verb|addarf| accordingly. In the end, we have six spectral sets for AGN in six flux bands.
%%%%%%%%%%%%%%%%%%%%%%%%%%%%%%%%%%%%%%%%%%%%--Spectral Fitting--%%%%%%%%%%%%%%%%%%%%%%%%%%%%%%%%%%%%%%%%%%%%%%%%%%%%%%%%%%
\section{Spectral Fitting}\label{Spectral Fitting}
To fully disentangle different components contributing to DXB, especially below 1 keV, as well as the contribution from the galaxies, we need to fit the spectrum sets in the 0.5-7.2 keV band. After the QPB is subtracted from the spectra, some instrumental backgrounds are still present. We excluded the energy band below 0.5 keV where the XMM signal-to-noise ratio drops very quickly and the low energy tails is exceptionally strong, as well as the strong Al K$\alpha$ instrumental line in 1.3 - 1.75 keV. The upper cutoff at 7.2 keV is to avoid more bright instrumental lines such as the Cu fluorescent instrumental line ($\sim$ 8 keV). 

\begin{figure}
\begin{center}
\includegraphics[width=8cm]{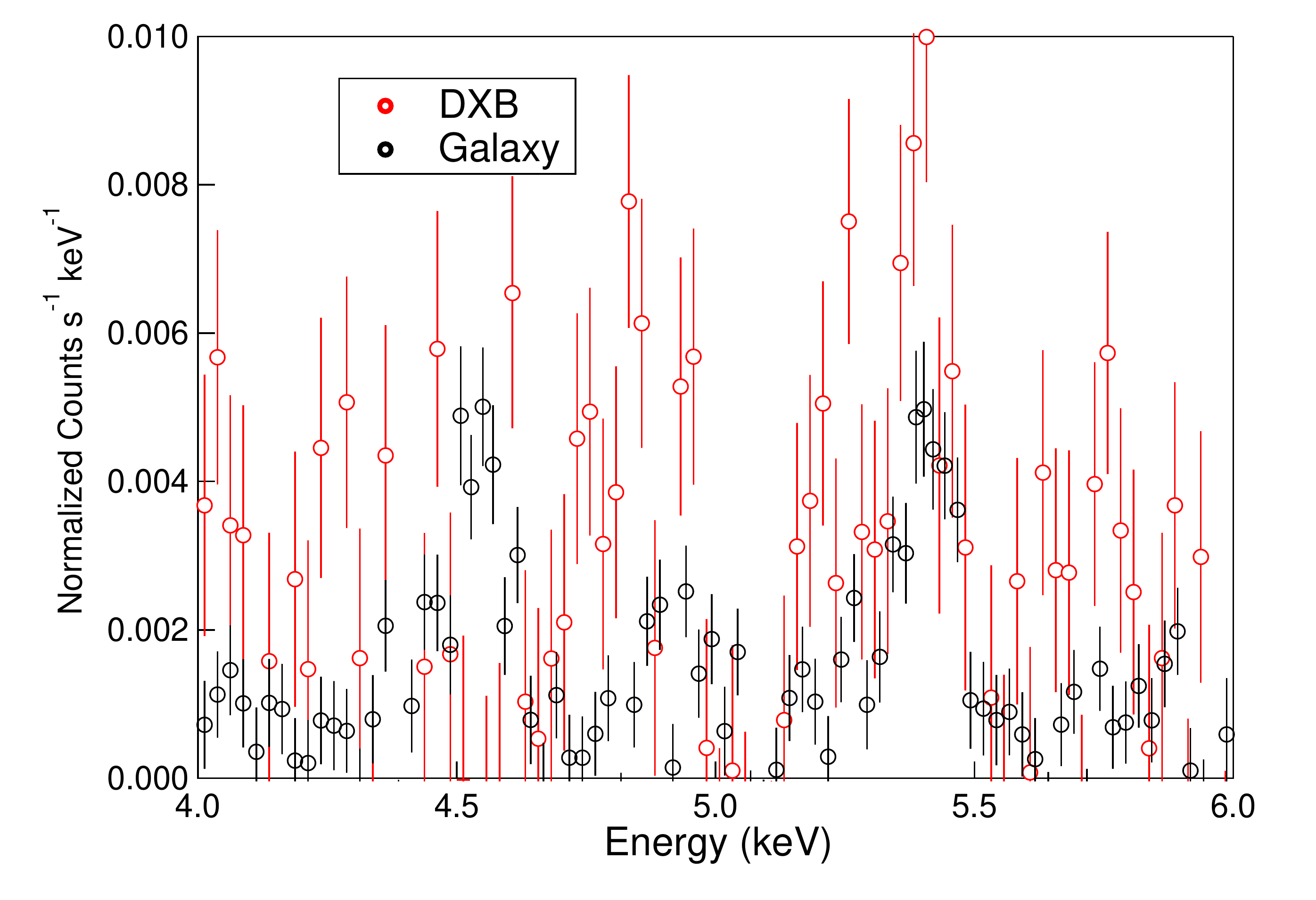}
\end{center}
\vspace{-0.2cm}
\caption{DXB and galaxy spectra in 4--6 keV. Two excess lines are apparent at 4.5 and 5.4 keV.}
\label{fig:4-6}
%\vspace{-12pt}
\end{figure}

Visual inspection of the galaxy spectra suggests that there are two emission lines present at 4.5 and 5.4 keV, 
which are less apparent but still visible in DXB spectra (see Figure \ref{fig:4-6}). They are identified as Ti and Cr emission lines, which are part of the XMM instrumental background emission lines \citep{XMMIns}. Even though those two lines were not as apparent in the DXB spectra, we also removed the 4.2-4.8 keV and 5.1-5.7 keV regions to make the fitting ranges uniform for all spectrum sets. 

As discussed in the introduction, all our data are affected by SWCX, which is time variable, thus not the same for all 32 pointings. For this reason, rather than merging the 32 spectra and fit the combined spectrum, we decided to perform a simultaneous fit of all 32 spectra. In the fit, all fitting parameters associated with non-varying components are linked together, while the parameters associated with SWCX are individual to each spectrum. Such procedure provides a better estimate of SWCX contamination to the astronomical sources and allows us to study the temporal variation of the SWCX emission. 

%%%%%%%%%%%%%%%%%%%%%%%%%%%%%%%%%%%%%%%%%%%%--DXB fitting--%%%%%%%%%%%%%%%%%%%%%%%%%%%%%%%%%%%%%%%%%%%%%%%%%%%%%%%%%%
\subsection{Measurement of the Diffuse X-ray Background}
Above 1 keV, emission from DXB is mostly due to unresolved point sources, while below 1 keV, the contribution is more complex. We fitted the 32 DXB spectra simultaneously using \verb|XSPEC| with a 5-component model: 
\verb|APEC+Gaussian+Gaussian+wabs(APEC+Powerlaw)|, where \verb|wabs| is the a photo-electric absorption using Wisconsin \citep{Morrison1983} cross-sections. 
Details of each component are discussed below.

\subsubsection{Local Hot Bubble} 
We modeled the emission of the LHB  using the thermal plasma model APEC. It is integrated into \verb|XSPEC| by using The Astrophysical Plasma Emission Code (APEC) from the AtomDB Project\footnote{\url{http://atomdb.org/}} to produces an emission spectrum from collisionally-ionized diffuse gas.   

Most of the LHB emission is below the energy range of XMM-Newton and only a small fraction contributes to our data. As such, instead of including the LHB in our fitting procedure, the temperature and emission measure are fixed to the values derived from the combined analysis of the ROSAT \& DXL (Diffuse X-rays from the Local Galaxy) missions \citep{Liu2017}.  For each observation, the normalization parameter of APEC can be then calculated by the emission measure and solid angle of the observation.

\subsubsection{Solar Wind Charge Exchange} 

When periods of high solar activity are avoided, most of the SWCX emission is in the 1/4 keV band and its contribution to higher energy is primarily limited to O VII and O VIII emission lines. Based on XMM-Newton energy resolution, these can be modeled by two Gaussian lines with zero line width at 0.57 keV and 0.65 keV.

As discussed before, to properly account for the temporal variation of SWCX emission,instead of combining all spectra from 32 observations into a single stacked spectrum and fitting it, we decided to keep the spectra separate and fit them simultaneously. 
This allows us to model the temporal variation of the SWCX in a 10-year span of the observations.
In our analysis, we also tested the ATOMDB Charge eXchange (ACX) model instead of two Gaussian lines \citep{Smith2014}, with the same fitting procedure, which gives us a similar result (see Section \ref{SWCX result}).

\subsubsection{Ne IX emission line} 

After fitting with the two-gaussian model, we noticed that there remains a strong excess around 0.9 keV, at the energy of Ne IX emission line. The He-like ion Ne IX is typically found in plasmas with a temperature range of $5\times10^5-10^7$K \citep{Mazzotta1998}, which can be originated from many types of astrophysical sources. To model for the Ne IX emission line, we add another Gaussian line between the energy range 0.8 keV and 1 keV. 

\subsubsection{Galactic Halo and Circum-galactic Medium} 
ROSAT all-sky survey and recent shadowing clouds study suggest that $\sim50\%$ of the 3/4 keV emission originates beyond the LHB \citep{Burrows1991, Snowden1991}, with a temperature $\sim 10^6 - 10^{6.5}$ K \citep{Snowden1998, Kuntz+Snowden2000, Smith2007, Galeazzi2007}. This emission is commonly thought of as a combined emission from Galactic Halo (GH) and CircumaGalactic Medium (CGM), and it can be modeled by an absorbed thermal plasma model \citep{Snowden1998} 
We fixed the neutral hydrogen column density $\textrm{N}_\textrm{H}$ to the average value in the CDFS field ($0.0088 \times 10^{22}$ cm$^{-2}$, \cite{Dickey1990}), and we used that to calculate the absorption factor with the wabs model\citep{Morrison1983}. 
For the thermal model APEC, the metal abundance is set to solar abundance and redshift set to zero. 

\subsubsection{Unresolved Point sources}
Point sources including AGN, normal and star-forming galaxies contribute most of the DXB above 1 keV. The spectrum of a point source is mainly due to synchrotron radiation and can be modeled by an absorbed powerlaw model \citep{Nandra1994, Reeves2000, Piconcelli2005}, with the same column density for the absorbed thermal component model. 
%%%%%%%%%%%%%%%%%%%%%%%%%%%%%%%%%%%%%%%%%%%%--GALAXIES fitting--%%%%%%%%%%%%%%%%%%%%%%%%%%%%%%%%%%%%%%%%%%%%%%%%%%%%%%%%%%

\subsection{Measurement of the  Galaxies}\label{Measuremnet of the Galaxies}

We fitted the spectra for galaxies from 32 observations simultaneously using the same models for DXB in 0.5-7.2 keV band. The normalization of the unabsorbed APEC is re-calculated and fixed according to the overall solid angles of galaxies in each observation. It is worth noting that, in principle, the spectrum is more complex. To begin with, there should be two absorbed thermal components, one from the GH/CGM and one from galaxies. Moreover, the population also includes highly redshifted galaxies, where the thermal component is outside our energy range and other mechamisms may be responsible for the detected emission, such as synchrotron/power-law component coming from the X-ray binary population in those galaxies.
However, as the spectra can be well fitted with a single thermal component, there is no statistical improvement in the use of more complex models. As our goal is primarily the evaluation of the contribution to the diffuse emission, a single absorbed thermal component approximation is therefore sufficient for our analysis. 
Moreover, we utilize the ratio of the thermal components normalization from all four sets of spectra to work out the relative contributions from GH/CGM and galaxies (see Section \ref{galaxy_vs_gh}).

%%%%%%%%%%%%%%%%%%%%%%%%%%%%%%%%%%%%%%%%%%%%--AGN fitting--%%%%%%%%%%%%%%%%%%%%%%%%%%%%%%%%%%%%%%%%%%%%%%%%%%%%%%%%%%
\subsection{Measurement of the AGN spectrum}\label{Measuremnet of the AGN}
Thanks to the large number of identified AGNs in the CDFS, we are able to categorize them into six flux groups. For each flux group, we fit the data with an absorbed powerlaw, an unabsorbed APEC model and an absorbed APEC model to account for a soft excess below 0.7 keV in some groups. The fitting was done separately for each flux group. Again, the hydrogen column density is set to $0.0088 \times 10^{22}$ cm$^{-2}$. The photon index and normalization of the powerlaw, as well as the parameters of the absorbed APEC, are left free to vary during the fitting. The temperature and emission measure of the APEC are fixed to values derived from ROSAT \& DXL, with normalization parameters scaled by the corresponding overall solid angles in each group.

%%%%%%%%%%%%%%%%%%%%%%%%%%%%%%%%%%%%%%%%%%%%--RESULTS--%%%%%%%%%%%%%%%%%%%%%%%%%%%%%%%%%%%%%%%%%%%%%%%%%%%%%%%%%%
%%%%%%%%%%%%%%%%%%%%%%%%%%%%%%%%%%%%%%%%%%%%--spectra--%%%%%%%%%%%%%%%%%%%%%%%%%%%%%%%%%%%%%%%%%%%%%%%%%%%%%%%%%%
\begin{figure*}
\begin{center}
\includegraphics[width=\textwidth]{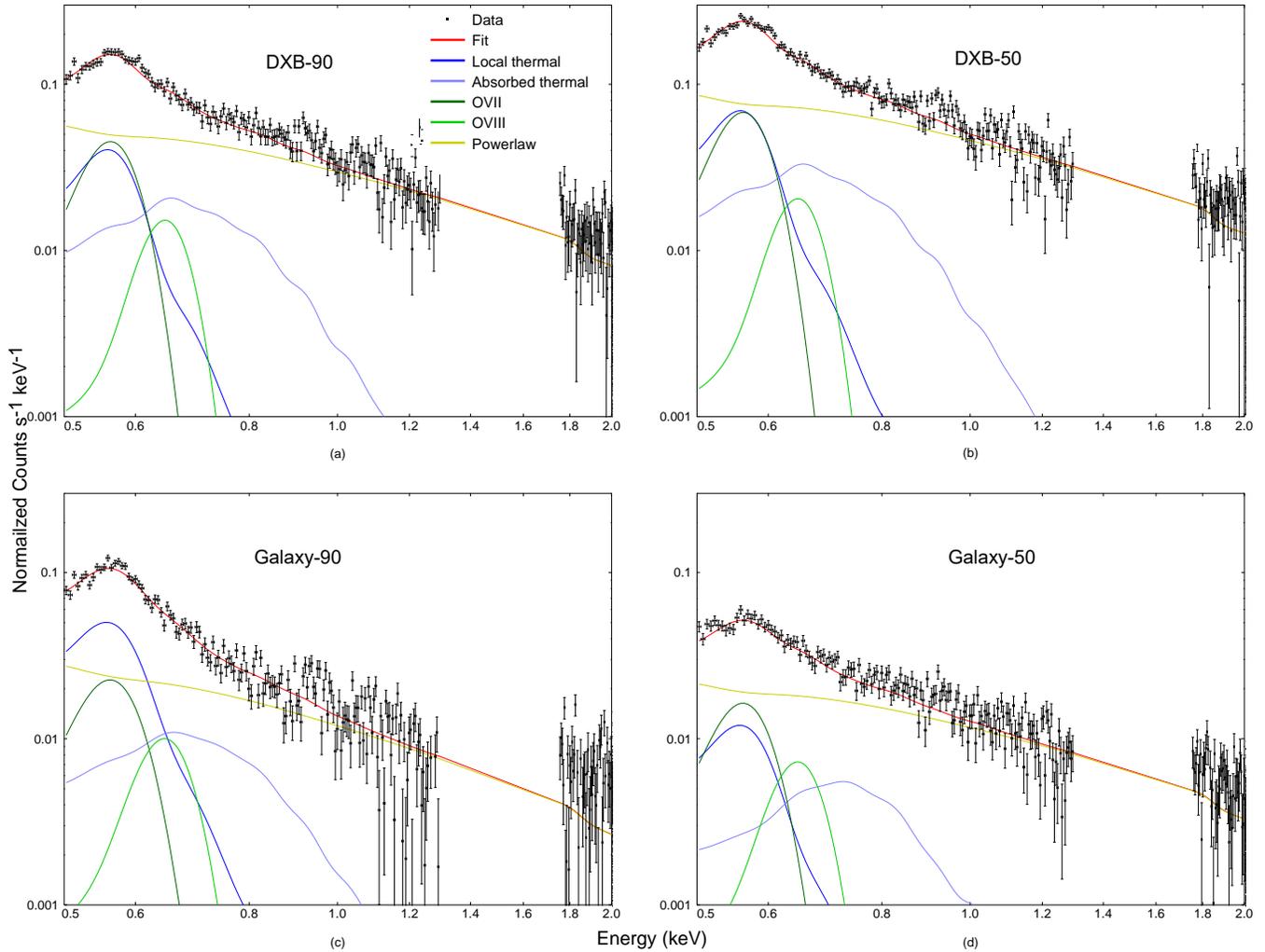}
\end{center}
\vspace{-0.2cm}
\caption{Cumulative spectra with fittings for: (a)DXB-90; (b)DXB-50; (c)Galaxy-90; (d)Galaxy-50}
\label{fig:spectra}
%\vspace{-12pt}
\end{figure*}
%%%%%%%%%%%%%%%%%%%%%%%%%%%%%%%%%%%%%%%%%%%%--spectra--%%%%%%%%%%%%%%%%%%%%%%%%%%%%%%%%%%%%%%%%%%%%%%%%%%%%%%%%%%

\section{Results}
While we performed a combined fit of all 32 datasets simultaneously, to visually compare our result with other works, composite source and background spectra of 32 observations are generated using \verb|mathpha|. Composite ARFs and RMFs are also generated by \verb|FTOOLS addarf| and \verb|addrmf|, respectively. The models in the plots are generated by a weighted average of the fitting parameters based on the exposure time of each observation. The composite spectra of DXB and galaxies in 0.5-2 keV band are shown in Figure \ref{fig:spectra}. All fitting parameters with uncertainties are listed in Table 2.

\subsection{Solar Wind Charge Exchange}\label{SWCX result}
%%%%%%%%%%%%%%%%%%%%%%%%%%%%%%%%%%%%%%%%%%%%--FIG--%%%%%%%%%%%%%%%%%%%%%%%%%%%%%%%%%%%%%%%%%%%%%%%%%%%%%%%%%%
\begin{figure*}
\begin{center}
\includegraphics[width=\textwidth]{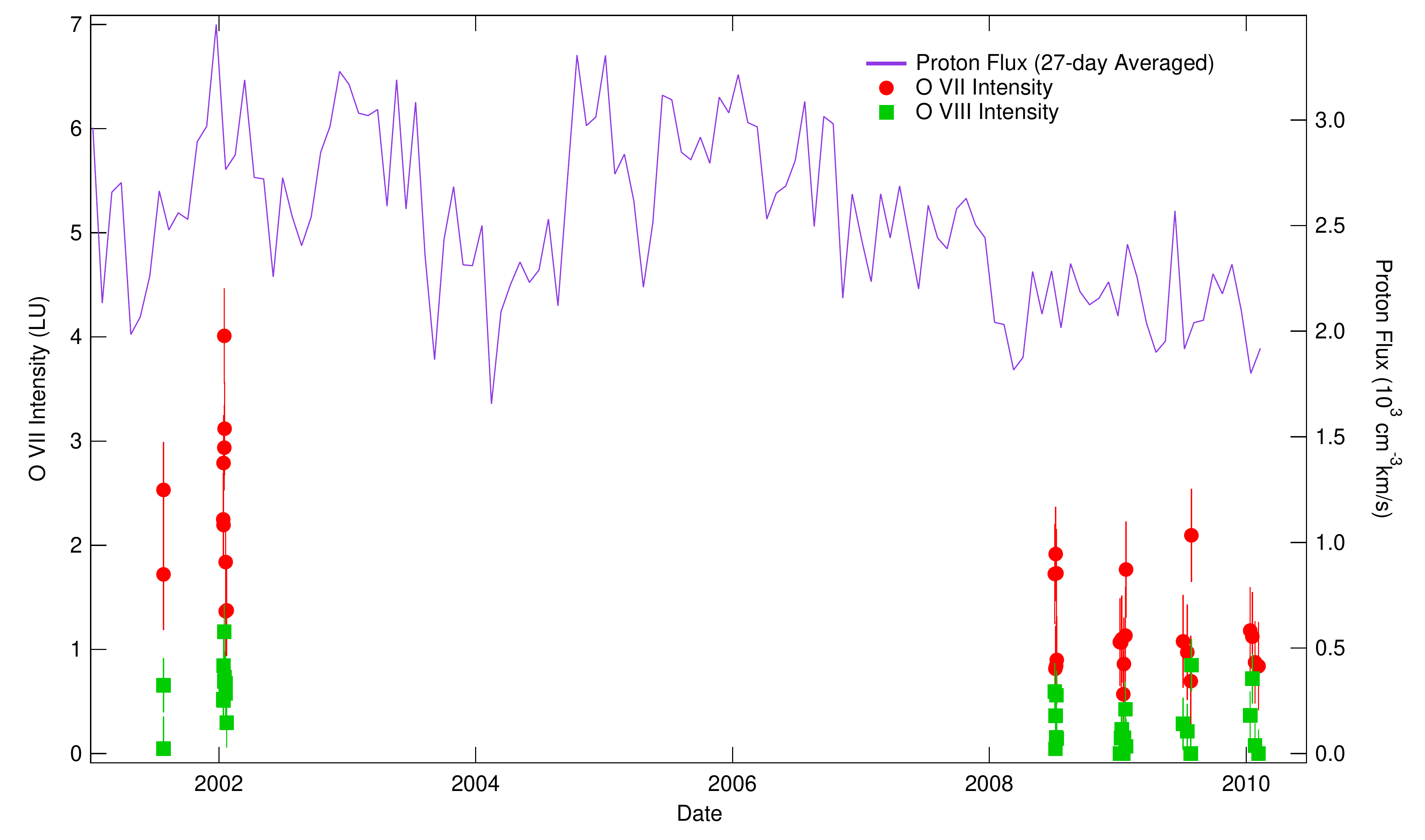}
\end{center}
\vspace{-0.2cm}
\caption{Correlation of O VII and O VIII intensities (LU) with solar proton flux}
\label{fig:O_trend}
\end{figure*}

\begin{figure}
\begin{center}
\includegraphics[width=7cm]{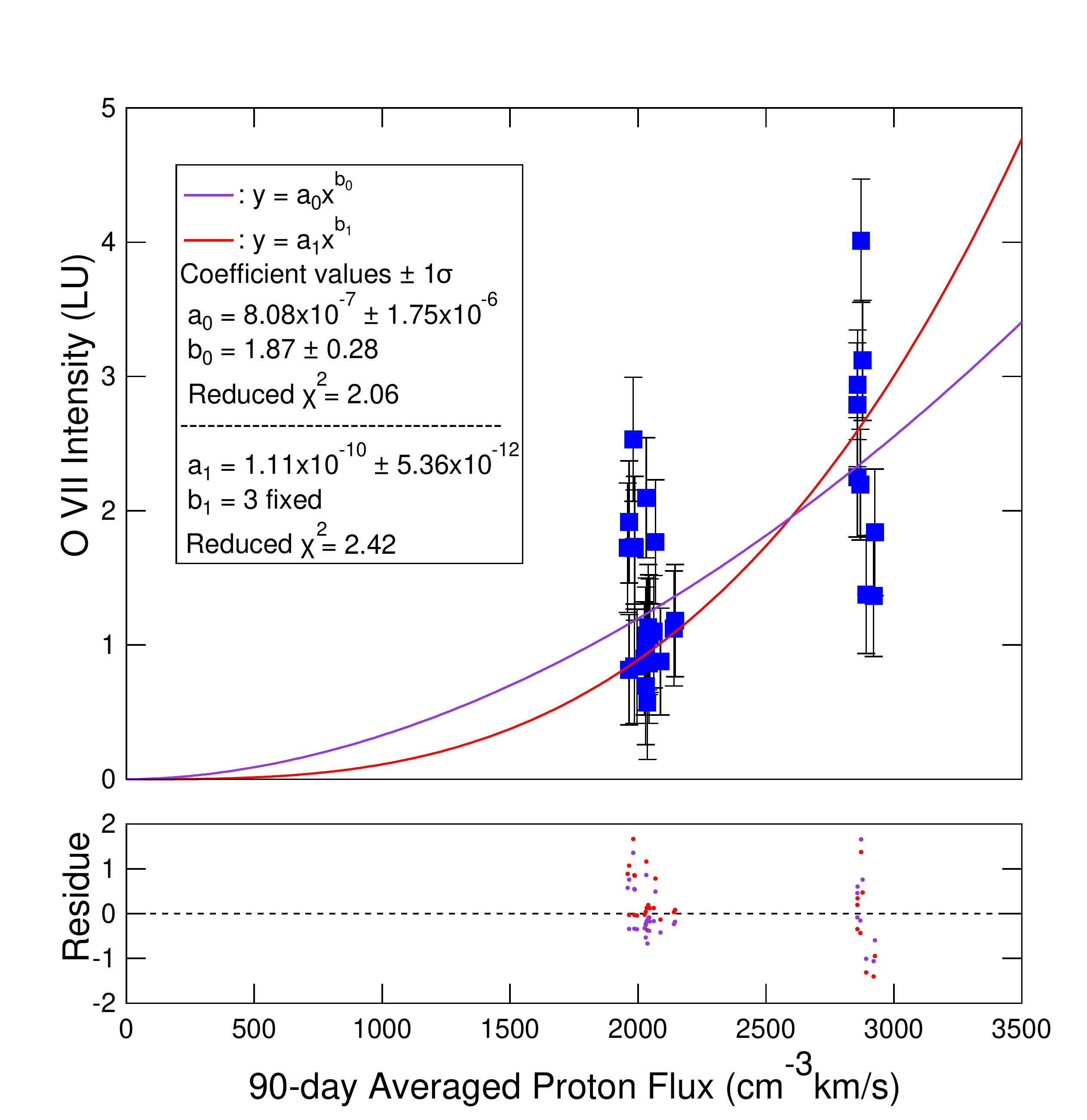}
\end{center}
\vspace{-0.2cm}
\caption{O VII intensity vs. 90-day averaged proton flux. The proton flux data is extracted from NASA/GSFC's OMNI data set through OMNIWeb service. The correlation between O VII emission and 90-day averaged proton flux can be described as a powerlaw function with the best-fit powerlaw index $\sim1.9$.}
\label{fig:O_pf_90}
\end{figure}

\begin{figure}
\begin{center}
\includegraphics[width=7cm]{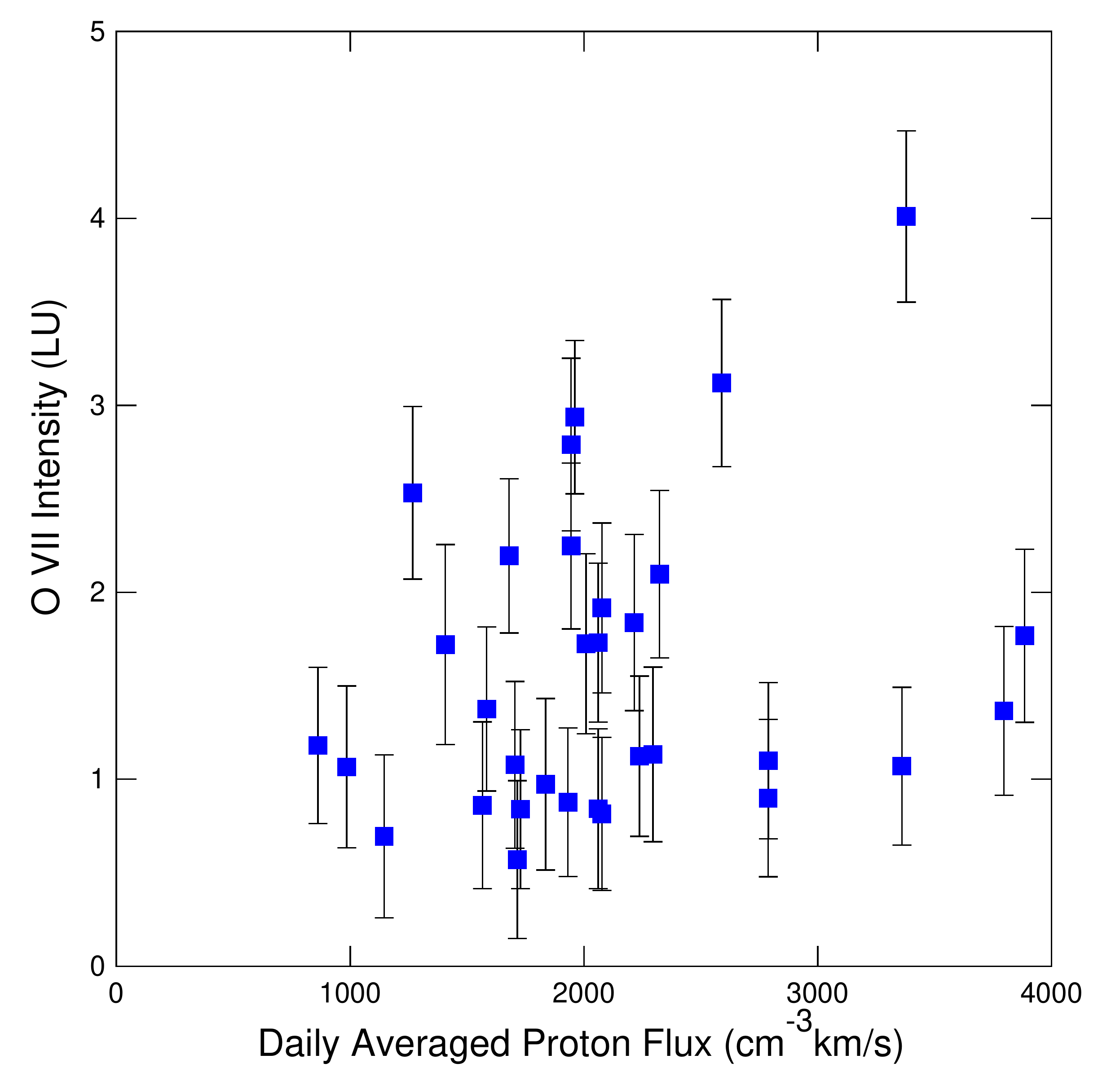}
\end{center}
\vspace{-0.2cm}
\caption{O VII intensity vs. daily averaged proton flux (from OMNI data). No significant relationship is found for O VII intensity and the daily averaged proton flux at the time of the X-ray pointing.}
\label{fig:O_pf}
\end{figure}
%%%%%%%%%%%%%%%%%%%%%%%%%%%%%%%%%%%%%%%%%%%%--FIG END--%%%%%%%%%%%%%%%%%%%%%%%%%%%%%%%%%%%%%%%%%%%%%%%%%%%%%%%%%%

The CDFS data were taken over the course of nine years from July 2001 through February 2010. Figure \ref{fig:O_trend} shows the variation of O VII and O VIII, together with solar proton flux during that time, showing a correlation between them. For example, the observation 0108061901 was taken during a period of solar maximum in 2002, and the O VII is as high as three times its average values during the solar cycle. The plot also shows the O VII emission to be as low as $\sim 1$ Line Unit (LU), setting an upper limit on the emission from GH/CGM. 

%%%%%%%%%%%%%%%%%%%%%%%%%%%%%%%%%%%%%%%%%%%%--FIG--%%%%%%%%%%%%%%%%%%%%%%%%%%%%%%%%%%%%%%%%%%%%%%%%%%%%%%%%%%
\begin{figure}
\begin{center}
\includegraphics[width=7cm]{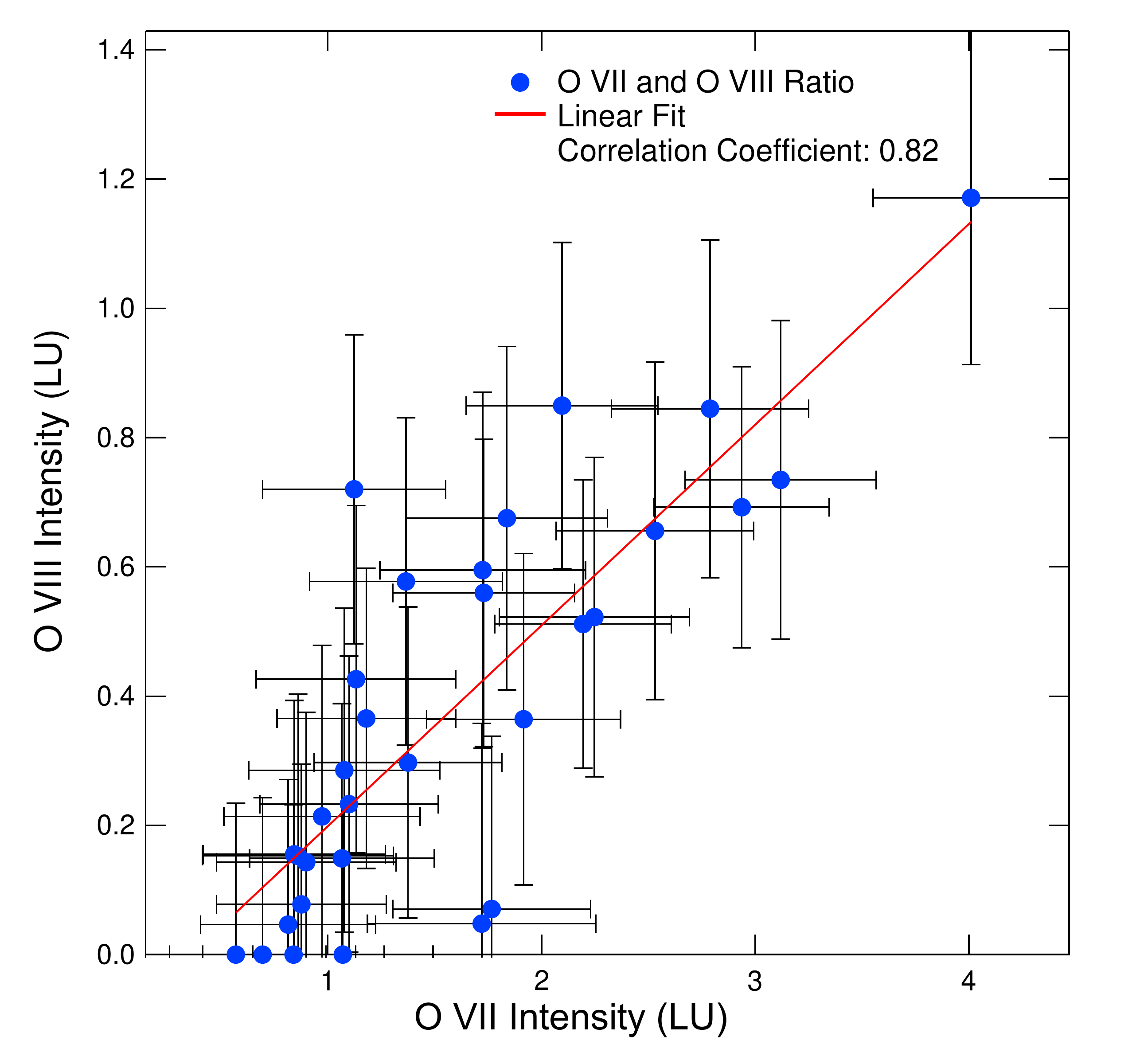}
\end{center}
\vspace{-0.2cm}
\caption{Linear fit to O VII and O VIII correlation}
\label{fig:O_ratio}
\end{figure}

\begin{figure}
\begin{center}
\includegraphics[width=7cm]{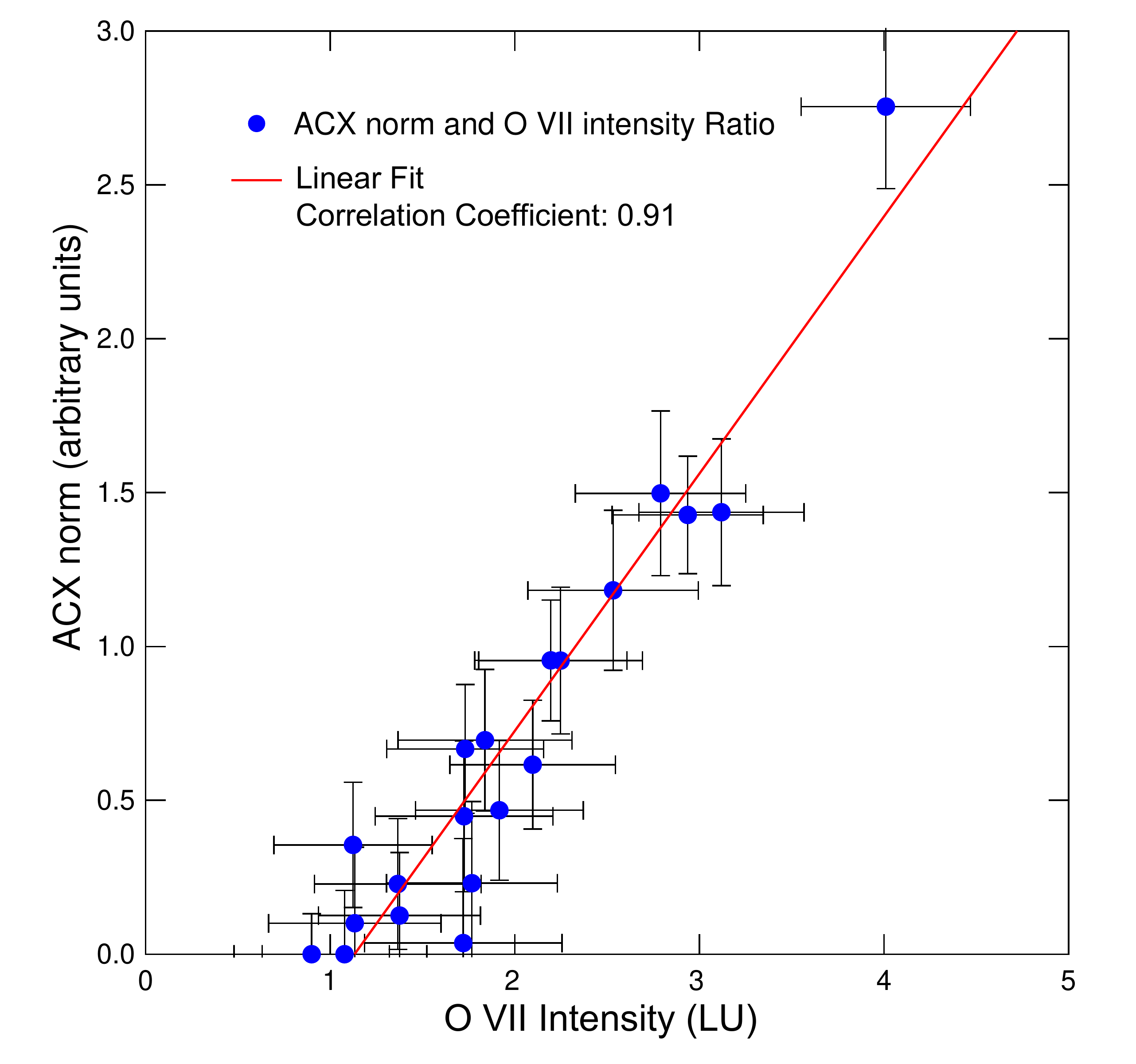}
\end{center}
\vspace{-0.2cm}
\caption{Variation of ACX norm vs. O VII intensity.}
\label{fig:acx_O_ratio}
\end{figure}
%%%%%%%%%%%%%%%%%%%%%%%%%%%%%%%%%%%%%%%%%%%%--FIG END--%%%%%%%%%%%%%%%%%%%%%%%%%%%%%%%%%%%%%%%%%%%%%%%%%%%%%%%%%%

Figure \ref{fig:O_pf_90} shows the relationship between O VII line emission and proton flux, showing a clear correlation between the two as a second power.
At first order, the line intensity is proportional to the flux of $O^{7+}$ in the solar wind. As $O^{7+}$ flux is not always readily available, we used the proton flux as a proxy. However, as determined by \cite{Liu2014HEAD}, the relationship between $O^{7+}$ flux and proton flux is not linear but cubic. Our result shows the power law fit performed with the software Origin (including error bars) with the best power law index of $1.87\pm0.28$. The discrepancy between our result and the prediction of \cite{Liu2014HEAD} can be due to the simplified mode, or the limited spread in proton flux of the data. We also tried to fix the power law index to 3 and the reduced $\chi^2$ increased 0.36 with 1 less degree of freedom. We note that, in the plot, following the recipe from \cite{Liu2014HEAD}, the proton flux is averaged over the 90 days before the XMM-Newton pointing, to account for the propagation of the solar wind through the interplatenary medium. We did not find any significant relationship if we only used the proton flux at the time of the X-ray pointing (Figure \ref{fig:O_pf}), as would have been the case for geocoronal SWCX. In addition to provide a recipe to study the relationship between O VII emission and proton flux, the results support a heliospheric origin for the emission. 

Figure \ref{fig:O_ratio} shows the correlation between O VII and O VIII. The relationship is linear, indicating that the ratio between $O^{7+}$ and $O^{8+}$ was, at first order, constant for all pointings.  The best fit parameter shows an O VIII/O VII ratio is $0.25\pm0.13$, with a correlation coefficient of 0.82. Our result is in agreement with previous results on O VIII/O VII studies such as \cite{Yoshino2009}, who derived an O VIII/O VII line intensity ratio of $\sim 0.22 - 0.67$. 

In addition to the simple 2-Gaussian model to describe the SWCX, we also used the ACX model, which is designed to approximate the likely contribution of Charge eXchange (CX) to the observed spectrum \citep{Smith2012}. This model assumes that each ion undergoes only a single CX in the line of sight and does not take into account the actual CX cross sections and relies on approximated estimates of the ions densities. As a result, it only provides a good estimate of the extent of the SWCX contamination. We fixed the SWCX parameters to their recommended values (see \cite{Smith2014}):
temperature = 0.85 MK 
, FracHe0 = 0.090909, abundance = Anders \& Grevesse solar values, redshift = 0, swcx = 1 and model = 8 (see \cite{Smith2014}) and fitted the 32 spectra simultaneously.
We used this model as a diagnostic tool to confirm the results from the two-Gaussian model. The two-Gaussian model can adequately explain and quantify the O VII and O VIII lines, which serve to constrain the average SWCX contamination accurately in CDFS over one solar cycle. 
Figure \ref{fig:acx_O_ratio} shows the correlation between ACX normalization and O VII intensity, with a correlation coefficient = 0.91, showing the strong correlation between the two procedures. In addition, Figure \ref{fig:acx_O_ratio} shows the existence of an intensity floor for O VII emission at $\sim 1$ LU. This can be explained by contribution from GH/CGM to the O VII emission.

\subsection{Ne IX line}

\begin{figure}
\begin{center}
\includegraphics[width=7cm]{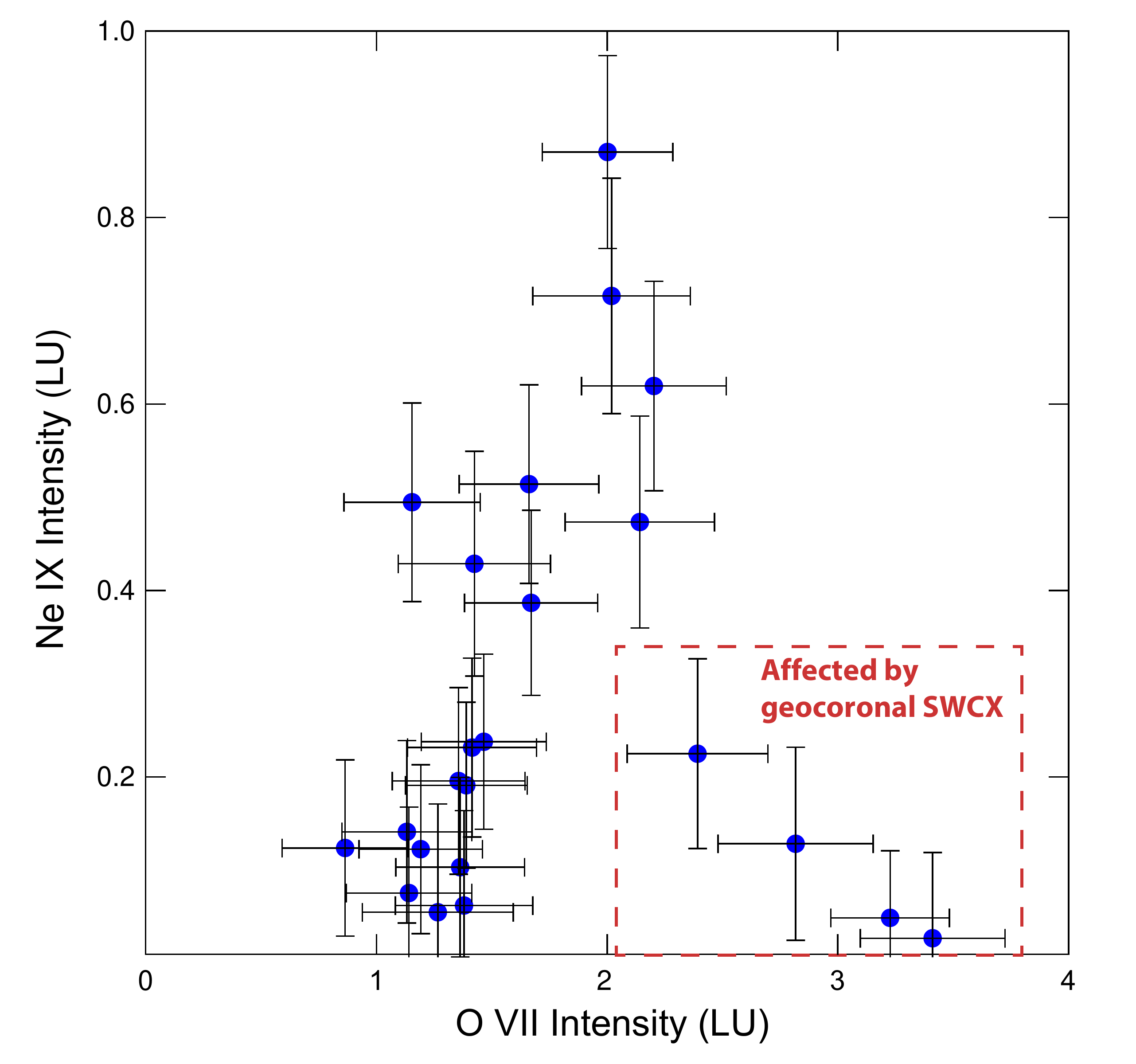}
\end{center}
\vspace{-0.2cm}
\caption{Ne XI intensity vs. O VII intensity.}
\label{fig:ne_o7}
\end{figure}

As mentioned before, we found an excess emission in our results consistent with the Ne IX emission line around 0.93 keV.
To study the properties of that emission, we added a Gaussian to our mode and decided to let the energy of the line vary in the fit to confirm its association with Ne IX. The parameters of the line shown in Table 4 are all consistent with 0.93 keV, confirming it is Ne IX. 
The comparison between multiple spectra shows that the line intensity is proportional to the solid angle and thus associated with diffuse emission, not the galaxy population. 
Moreover, Figure \ref{fig:ne_o7} shows the relationship between the Ne IX intensity and the O VII intensity. The figure indicates a positive correlation between Ne IX and O VII emissions, except for the four datapoints with the highest O VII intensities. This suggest that part of the Ne IX emission is originated by heliospheric SWCX.
For the four datapoints that do not seem to follow this trend, we looked at solar wind conditions and two of them are clearly related to a spike in proton flux at the time of the observation, leading to the belief that they are additionally affected by Geocoronal SWCX emission. While a spike is not as obvious for the remaining two datapoints, we assume the same origin. 

\subsection{Emission from Galaxies and Galactic Halo/Circum-Galactic Medium}\label{galaxy_vs_gh}

The emission from GH/CGM is typically associated with an absorbed thermal component with a temperature of $\sim$3~MK. However, emission from galaxies is also expected contains a significant thermal component with similar characteristics, and while galaxies may be redshifted, due to the limited energy resolution of XMM-Newton, the effect is limited. As a consequence, we decided that the two components cannot be analyzed independently of each other. Following the recipe discussed in section \ref{Spectra generation}, we fit each spectrum independently with a model containing a single absorbed thermal component, and the results are shown in Table 2. 

Before we combine them, we note that the temperatures of the absorbed thermal component in the galaxy spectrum sets (0.3 and 0.33 keV) are typically higher than the temperature of DXB absorbed thermal component (0.28 and 0.27 keV), indicating that the temperature of GH/CGM can generally be lower than normal galaxies in the direction of CDFS. 
We also found that our measured temperature is in general higher than previous work \cite{Galeazzi2007, McCammon2002} while the emission measure is smaller.
While the different emission measure is likely due to anisotropy in the emission over the sky, the difference in temperature may be due to the fact that previous work did not include the SWCX in their fit. As such, the temperature is affected by an excess oxygen emission, which would have the net effect of lowering the temperature. 

Table 3 shows the absorbed thermal component normalization, together with the total solid angles and fluxes in 0.5-7 keV band for all 4 spectra. For the DXB spectra, the normalization has also been converted into the Emission Measure. 
To separate the GH/CGM contribution from that of galaxies, we use the fact that, due to our data selection, vignetting effects are negligible and, at first order, the normalization of diffuse emission (GH/CGM) scales linearly with the solid angle, while the normalization of point sources (galaxies) scales as the PSF. For example, the normalization between Galaxy-50 and Galaxy-90 should change by a factor of 7.61/1.59=4.8 if due to GH/CGM, or by a factor of 90\%/50\%=1.8 if due to galaxies. 
The results from Table 3 shows that the component in the galaxy spectra is probably a combination of both diffuse and point source, although closer to point source. We then set up a simple system of equations assuming that both a diffuse and point source component affect both Galaxy-50 and Galaxy-90. Assuming that the ratio of the diffuse component scales as 4.8 between the two datasets, while the point source component scales as 1.8, the system of equations can be solved to find each components separately. 
What we found is that $\sim57\%-78\%$ of the flux of the absorbed thermal component in the galaxy spectra originates from the galaxies themselves, and only $\sim 22\%-43\%$ of the flux comes from the foreground emission due to GH/CGM.

Considering the fact that most of the galaxies used for this work are to faint to be detected in typical observations, 
we calculated their contribution to the diffuse absorbed thermal emission typically associated with GH/CGM. In practice, if these galaxies are too faint to be resolved, their thermal emission would add to the diffuse emission. To estimate this contribution, we ``spread'' the emission over the all field of view and accounted for both photons that are outside the selected portion of the PSF and for galaxies that are event fainter than the Chandra sensitivity used here. 

\begin{figure}
\begin{center}
\includegraphics[width=7cm]{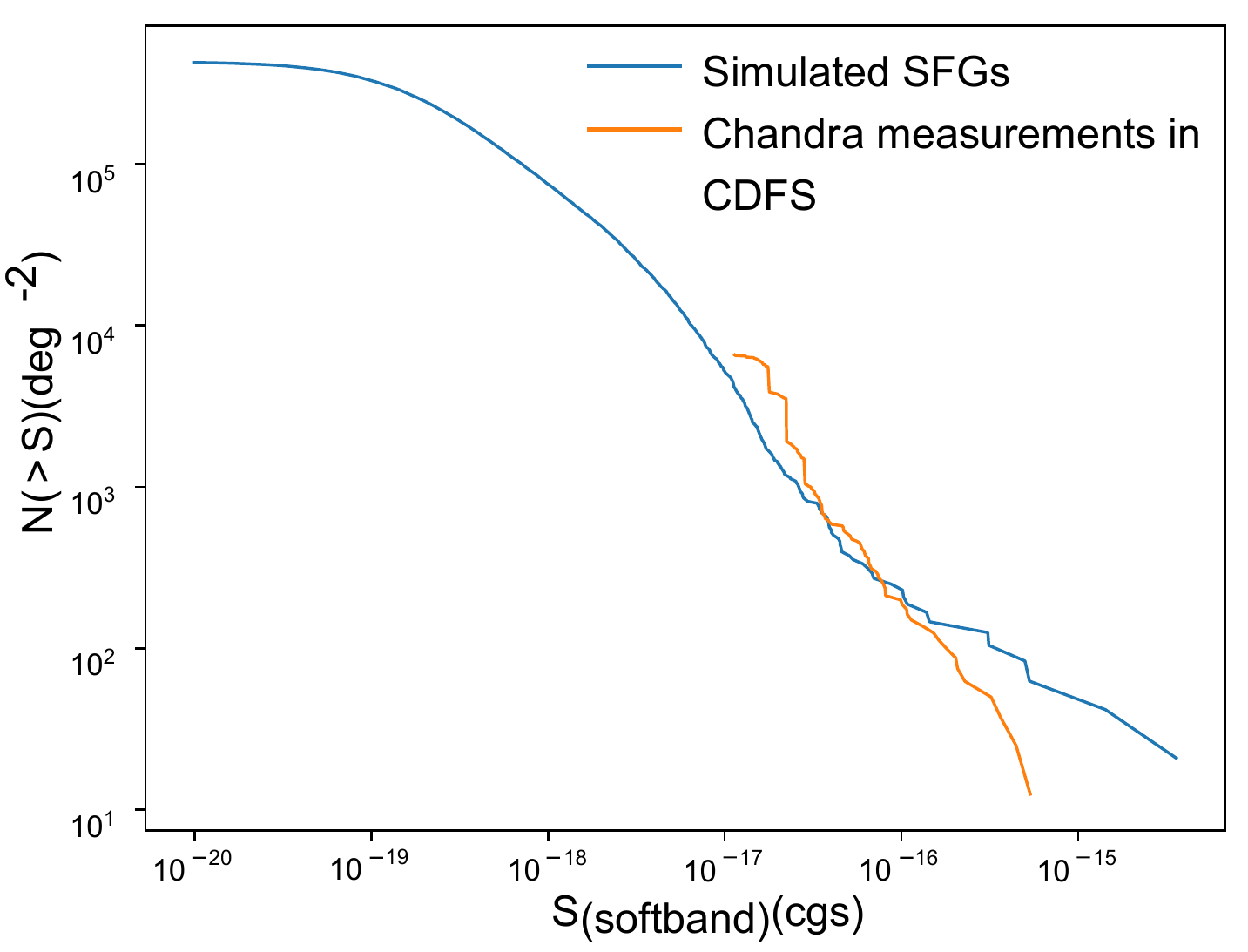}
\end{center}
\vspace{-0.2cm}
\caption{Camparison of the log$N$-log$S$ of simulated [0.5-2] kev SFGs in CANDELS with Chandra-detected galaxies in CDFS.}
\label{fig:logNlogS}
%\vspace{-12pt}
\end{figure}

Even though with Chandra's superb angular resolution, some galaxies are still too faint to be resolved. 
Evidence shows that below Chandra flux threshold, star-bursting and normal galaxies may become dominant in numbers \citep{Hickox2006}. \cite{Nico2016} created a sample of simulated star-forming galaxies(SFGs) in the Cosmic Assembly Near-infrared Deep Extragalactic Legacy Survey (CANDELS) field \citep{Grogin2011} based on estimated Infrared luminosities, which are calculated from their observed photometric redshift, stellar mass, UVJ rest-frame colors, and observed (or extrapolated) UV luminosities. Figure \ref{fig:logNlogS} shows the cumulative number counts log$N$-log$S$ for simulated SFGs in CANDELS and Chandra detected galaxies in CDFS. As we can see, the cumulative numbers of Chandra detected galaxies used in this work follow a similar trend as the simulated SFGs in CANDELS in the region covered by both, except for a steeper drop at the higher flux end due to the limited sample in our work. Extrapolating from the logN-logS plot and the portion of the flux outside the radius used, 
the total EM of unresolved galaxies should be $\sim3.36$ larger than the total EM of Chandra detected galaxies. 

%%%%%%%%%%%%%%%%%%%%%%%%%%%%%%%%%%%%%%%%%%%%--Figure 8--%%%%%%%%%%%%%%%%%%%%%%%%%%%%%%%%%%%%%%%%%%%%%%%%%%%%%%%%%%
%%%%%%%%%%%%%%%%%%%%%%%%%%%%%%%%%%%%%%%%%%%%--Figure 8 END--%%%%%%%%%%%%%%%%%%%%%%%%%%%%%%%%%%%%%%%%%%%%%%%%%%%%%%%%%%
To compute the total contribution from galaxies to the diffuse emission, we therefore calculated the contribution from resolved galaxies in the CFDS and spread it through the whole solid angle of the observation. This gives us an ``Emission Measure'' due to detected galaxies ($EM_{CDFS}$). We then multiply this value by 3.36 to get the total contribution from galaxies ($EM_g$). 
and multiply it by 3.36 to include contribution from unresolved galaxies based on \citep{Nico2016}:
\begin{equation} \label{eq:EM}
EM_g = EM_{CDFS} \times 3.36. %\frac{\Omega_g}{\Omega_g+\Omega_{dxb}} \times 3.36
\end{equation}
The resulting emission from galaxies (which are unresolved in typical observations) is $1.7\times10^{-4}$ cm$^{-6}$pc. 
In the CDFS direction, this is consistent with most of the ``diffuse'' emission being actually due to galaxies and not GH/CGM. This is not necessarily unexpected, as the absorbed thermal emission is know to be low in this direction and there is evidence of other fields where the absorbed thermal components is consistent with zero \citep{Ursino2014}. 

 To look at the more general effect this added galaxy component may have, \cite{Gupta2012} derived the electron density and path-length of the CGM  based on its emission measure. In their work, they used an average emission measure of $3\times10^{-3}$ cm$^{-6}$pc, with the electron density $n_e = 2.0\pm0.6\times10^{-4}$ cm$^{-3}$, path-length $L > 139$ kpc, and total baryonic mass $M_{total} = (2.3\pm2.1)\times10^{11} M_{\odot}$. If we consider that the unresolved galaxies contribution to the emission measure of GH is around $1.7\times10^{4}$ cm$^{-6}$pc, the resultant emission measure of GH will be $2.83\times10^{-3}$ cm$^{-6}$pc, and the corresponding parameters will be $n_e = 1.89\pm0.57\times10^{-4}$ cm$^{-3}$, $L > 147$ kpc, and $M_{total} = (2.6\pm2.4)\times10^{11} M_{\odot}$.

\subsection{AGN}
%%%%%%%%%%%%%%%%%%%%%%%%%%%%%%%%%%%%%%%%%%%%--Figure 6--%%%%%%%%%%%%%%%%%%%%%%%%%%%%%%%%%%%%%%%%%%%%%%%%%%%%%%%%%%
\begin{figure*}
\begin{center}
\includegraphics[width=\textwidth]{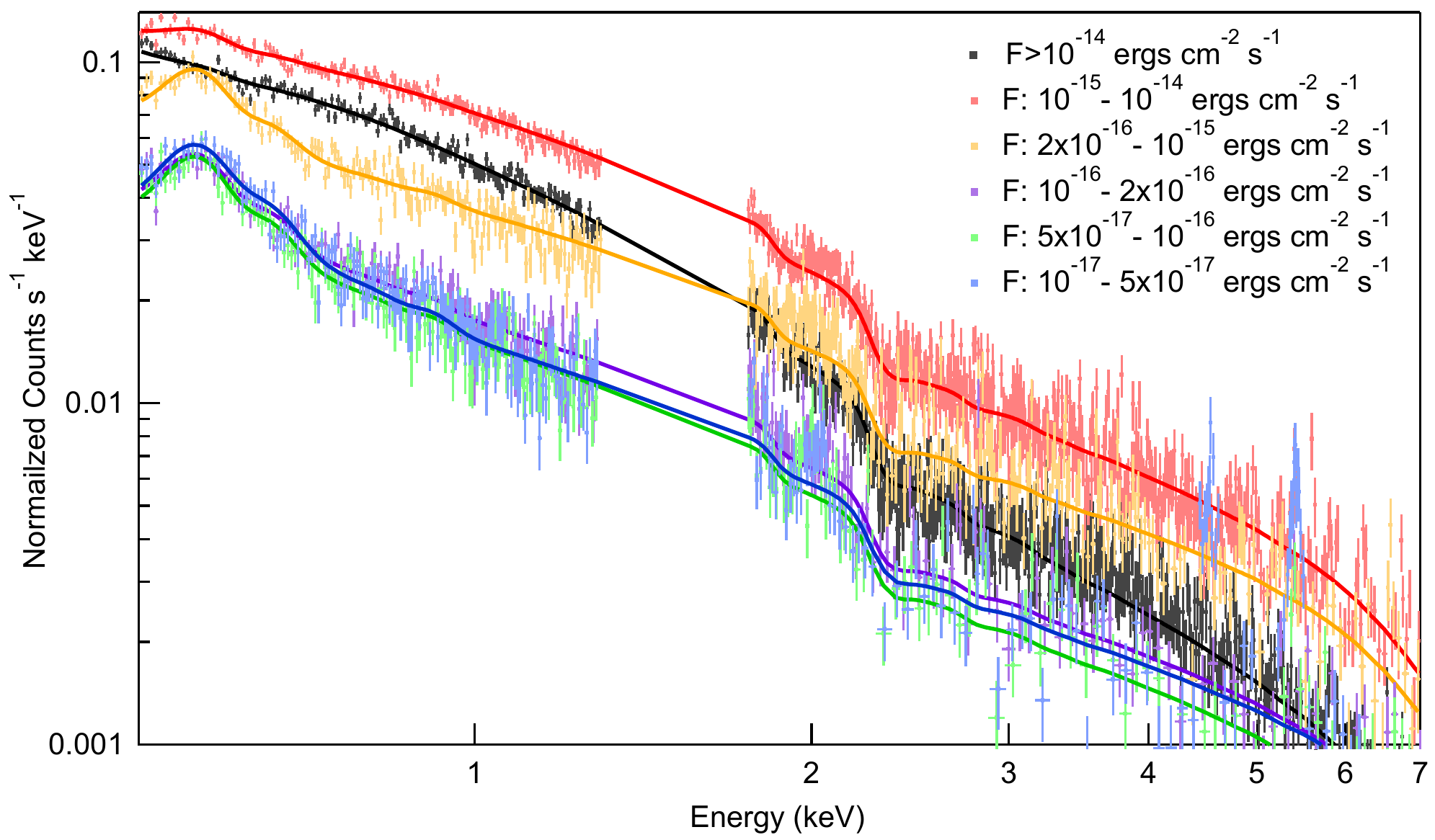}
\end{center}
\vspace{-0.2cm}
\caption{AGN Spectra in flux groups}
\label{fig:agn_spec}
\end{figure*}
%%%%%%%%%%%%%%%%%%%%%%%%%%%%%%%%%%%%%%%%%%%%--Figure 6 END--%%%%%%%%%%%%%%%%%%%%%%%%%%%%%%%%%%%%%%%%%%%%%%%%%%%%%%%%%%
%%%%%%%%%%%%%%%%%%%%%%%%%%%%%%%%%%%%%%%%%%%%--Figure 6--%%%%%%%%%%%%%%%%%%%%%%%%%%%%%%%%%%%%%%%%%%%%%%%%%%%%%%%%%%
\begin{figure}
\begin{center}
\includegraphics[width=7cm]{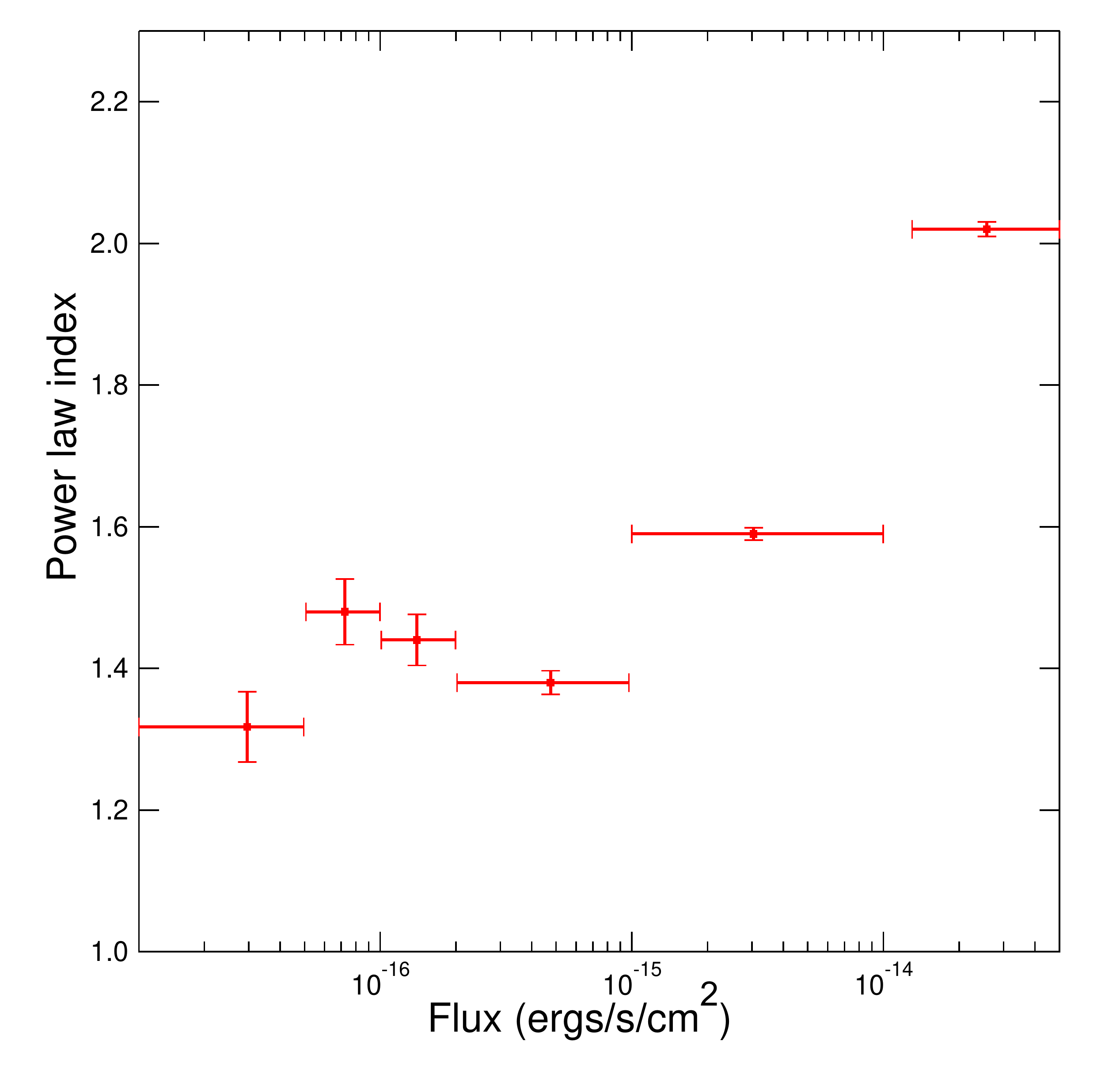}
\end{center}
\vspace{-0.2cm}
\caption{Variation of AGN photon index with flux}
\label{fig:p_index}
\end{figure}
%%%%%%%%%%%%%%%%%%%%%%%%%%%%%%%%%%%%%%%%%%%%--Figure 6 END--%%%%%%%%%%%%%%%%%%%%%%%%%%%%%%%%%%%%%%%%%%%%%%%%%%%%%%%%%%

As part of this investigation, we also looked at the contribution by AGN to the diffuse X-ray emission. While plenty of work has been dedicated to understand AGN in the CDFS, our goal was to quantify their contribution to the diffuse emission, particularly below 1 keV. 
In most observations, the
flux threshold for point source detection is high and most AGN remain unresolved. Our primary focus here was to see if and how the cumulative spectrum changes and whether extending the powerlaw fit from high energy to low energy, as it is commonly done, is appropriate. 

Figure \ref{fig:agn_spec} shows the spectral fittings for AGN categorized by their fluxes. The results are also summarized in Table~5. Referring to our brightness selected groups (section \ref{AGN Spectra creation}), we did not see any significant change from a single power law throughout almost all of the energy range (0.5-7 keV), confirming the robustness of our approach. We did see a small excess at the low end of the energy interval, consistent with thermal emission, but that was due to foreground contamination. 
That emission is consistent with zero in the brightest group and becomes more prominent for the fainter groups. This is consistent with thermal foreground.

In Fig.~\ref{fig:p_index} we also looked at how the power law index depends on AGN brightness. This is important for diffuse studies, as different investigation may have different point source sensitivity and the same power law index cannot be used. Overall, the index changes from about 2 for the brightest AGN, to about 1.3 for the faintest ones. This is not necessarily surprising, as it is known that faint AGN tend to have a harder spectrum  \citep{Gilli2007,Hasinger2008,Ananna2019}. We also note that the change is not fully monotonic, but we cannot say for sure whether that is just due to statistical fluctuations or some real physical properties of the sources.

\section{Conclusion}

In this work, we investigated the contributions of different components to the Diffuse X-ray Background in the direction of CDFS. We made use of 32 XMM-Newton observations of CDFS with a nominal exposure time $\sim 3.5$ Ms to study the spectral features of Diffuse X-ray Background in the energy range 0.5-7.2 keV. Here we summarize our main results:
\begin{enumerate}
  \item We confirm the correlation between O VII and O VIII emission line intensities with the O VIII/O VII ratio $=0.25\pm0.13$, whihc is in agreement with previous results from \cite{Yoshino2009}. We also find that the O VII emission is correlated with the 90-day averaged proton flux as a second power. In addition, by comparing the results from the two-Gaussian model and ACX model, we find the existence of an intensity floor for O VII emission at $\sim 1$ LU, also consistent with \cite{Yoshino2009}. This result supports the theory that the O VII emission arises from both the local SWCX+LHB and the more distant hot gas from GH/CGM, and the O VII emission from GH/CGM can be quantified as $\sim 1$ LU in the direction of CDFS. Moreover, by studying the relationship between the Ne IX intensity and the O VII intensity, we discover a positive correlation between Ne IX and O VII emissions, suggesting that part of the Ne IX emission is originated by heliospheric SWCX.
  \item By comparing the spectra of DXB and unresolved galaxies by XMM, we note that the temperatures of absorbed \verb|apec| component from galaxies are typically higher than the temperature of DXB absorbed thermal component, indicating that the temperature of GH/CGM can generally be lower than normal Galaxies in the direction of CDFS. 
  We calculate the emission measure from unresolved galaxies ($1.7\times10^{-4}$ cm$^{-6}$pc), while using simulated galaxies from \cite{Nico2016} to account for the total emission from galaxies below XMM and Chandra flux threshold. We find that the unresolved galaxies can contribute 26-68\% of the GH/CGM emission in CDFS, and $\sim 6\%$ of the all-sky average value of GH/CGM emission ($3\times10^{-4}$ cm$^{-6}$pc). For observations of GH/CGM with limited exposure times to resolve faint point sources, our result can serve as a baseline to remove the contribution from unresolved galaxies to the GH/CGM emission, and constrain the GH/CGM parameters such as electron density and pathlength more precisely.
  \item To study the relationship of AGN powerlaw index and flux, we divide Chandra-detected AGN in CDFS into six groups based on their fluxes. We find that the variation of AGN powerlaw index with flux generally follows a positive trend with power law index increasing as flux increases, with exceptions for two groups with low fluxes. For those two groups, the number of obscured AGN decreases while the powerlaw index increases. This suggests that obscuration alone cannot explain the variation in photon index. We also discover the spectrum for AGN with lower fluxes exhibits an excess below 1 keV, indicating that fainter AGN need to be modeled by an additional thermal component in addition to a simple powerlaw. 

\end{enumerate}

\acknowledgments

This research has made use of software provided by the European Space Agency in the application package XMM-ESAS for the analysis of pn observations. The scientific results reported in this article are based on observations made by the XMM-Newton X-ray Observatory, and data obtained from the Chandra Data Archive. 

%%%%%%%%%%%%%%%%%%%%%%%%%%%%%%%%%%%%%%%%%%%%--BIBLIOGRAPHY--%%%%%%%%%%%%%%%%%%%%%%%%%%%%%%%%%%%%%%%%%%%%%%%%%%%%%%%%%%

%%%%%%%%%%%%%%%%%%%%%%%%%%%%%%%%%%%%%%%%%%%%--TABLE 1--%%%%%%%%%%%%%%%%%%%%%%%%%%%%%%%%%%%%%%%%%%%%%%%%%%%%%%%%%%
\begin{deluxetable*}{cccccc}
\centering
\tablecolumns{6}
\tablewidth{0pt}
\tablecaption{Summery of XMM Observations for CDFS Used in This Investigation \tablenotemark{a}}
\label{tab:osbservations}
\tablehead{
\colhead{Obs.ID} & \colhead{RA} & \colhead{DEC} & \colhead{Start Date} & \colhead{End Date} & \colhead{Duration (s)}}
\startdata
0108060401 & 53.11125 & -27.805556 & 2001-07-27 & 2001-07-27 & 49906 \\
0108060501 & 53.122083 & -27.811111 & 2001-07-27 & 2001-07-28 & 64267 \\
0108060601 & 53.116667 & -27.813889 & 2002-01-13 & 2002-01-14 & 65318 \\
0108060701 & 53.11125 & -27.811111 & 2002-01-14 & 2002-01-15 & 94021 \\
0108061801 & 53.116665 & -27.80833 & 2002-01-16 & 2002-01-17 & 63018 \\
0108061901 & 53.116667 & -27.802778 & 2002-01-17 & 2002-01-18 & 54218 \\
0108062101 & 53.12208 & -27.80556 & 2002-01-20 & 2002-01-21 & 62119 \\
0108062301 & 53.116667 & -27.802778 & 2002-01-23 & 2002-01-24 & 88620 \\
0555780101 & 53.176245 & -27.75139 & 2008-07-05 & 2008-07-06 & 133118 \\
0555780201 & 53.176245 & -27.75972 & 2008-07-07 & 2008-07-08 & 133416 \\
0555780301 & 53.16666 & -27.75972 & 2008-07-09 & 2008-07-10 & 123811 \\
0555780401 & 53.16666 & -27.75139 & 2008-07-11 & 2008-07-12 & 122843 \\
0555780501 & 53.10417 & -27.82361 & 2009-01-06 & 2009-01-08 & 113004 \\
0555780601 & 53.10417 & -27.83195 & 2009-01-10 & 2009-01-12 & 118413 \\
0555780701 & 53.10417 & -27.84028 & 2009-01-12 & 2009-01-14 & 118415 \\
0555780801 & 53.094585 & -27.82361 & 2009-01-16 & 2009-01-18 & 120919 \\
0555780901 & 53.094585 & -27.83195 & 2009-01-18 & 2009-01-20 & 121518 \\
0555781001 & 53.094585 & -27.84028 & 2009-01-22 & 2009-01-24 & 125813 \\
0555782301 & 53.094585 & -27.84028 & 2009-01-24 & 2009-01-26 & 125714 \\
0604960101 & 53.176245 & -27.75972 & 2009-07-27 & 2009-07-28 & 129439 \\
0604960201 & 53.16666 & -27.75972 & 2009-07-17 & 2009-07-18 & 121094 \\
0604960301 & 53.176245 & -27.76917 & 2009-07-05 & 2009-07-06 & 122302 \\
0604960401 & 53.16666 & -27.76861 & 2009-07-29 & 2009-07-30 & 133915 \\
0604960501 & 53.10417 & -27.82361 & 2010-01-18 & 2010-01-19 & 46983 \\
0604960601 & 53.10417 & -27.83195 & 2010-01-26 & 2010-01-28 & 125212 \\
0604960701 & 53.094585 & -27.82361 & 2010-01-12 & 2010-01-14 & 120819 \\
0604960801 & 53.10417 & -27.84194 & 2010-02-05 & 2010-02-07 & 121087 \\
0604960901 & 53.094165 & -27.84297 & 2010-02-11 & 2010-02-13 & 125344 \\
0604961001 & 53.094585 & -27.83195 & 2010-02-13 & 2010-02-15 & 122515 \\
0604961101 & 53.10417 & -27.81453 & 2010-01-04 & 2010-01-06 & 120817 \\
0604961201 & 53.09292 & -27.81453 & 2010-01-08 & 2010-01-10 & 120718 \\
0604961801 & 53.094585 & -27.83195 & 2010-02-17 & 2010-02-19 & 125042 \\
\enddata
\tablenotetext{a}{Data from XMM-Newton Science Archive}
\end{deluxetable*}
%%%%%%%%%%%%%%%%%%%%%%%%%%%%%%%%%%%%%%%%%%%%--TABLE 2--%%%%%%%%%%%%%%%%%%%%%%%%%%%%%%%%%%%%%%%%%%%%%%%%%%%%%%%%%%
%\begin{longrotatetable}
\begin{deluxetable*}{lcccccc}[ht]
\tablecaption{Model parameter of fits to DXB and Galaxy spectra}
\label{tab:par}
\rotate
\tablehead{
\colhead{Parameter} & \colhead{DXB-90} & \colhead{DXB-50} & \colhead{Galaxy-90} & \colhead{Galaxy-50} & \colhead{Galeazzi\tablenotemark{a}} & \colhead{McCammon}\tablenotemark{b}}
\startdata
\sidehead{\textbf{Unabsorbed plasma component:}}
Temperature(keV) & 0.099 & 0.099 & 0.099 & 0.099 & 0.095 & 0.099  \\
Temperature($\times10^6$ Kelvin) & $1.15$ & $1.15$ & $1.15$ & $1.15$ & $1.11$ & $1.15$ \\
Emission Measure (cm$^{-6}$pc) & 0.0036 & 0.0036 & 0.0036 & 0.0036 & 0.0082 & 0.0088 \\ \tableline
%%\\tablebreak
\sidehead{\textbf{Absorbed plasma component:}}
Temperature(keV)& $0.28 \pm 0.0045$ & $0.27 \pm 0.047$ & $0.30 \pm 0.080$ & $0.33 \pm 0.057$ & 0.19 & 0.23 \\
Temperature($\times10^6$ Kelvin) & $3.25 \pm 0.052$ & $3.13 \pm 0.55$  & $3.48 \pm 0.93$ & $3.83 \pm 0.66$ & $2.27$ & $2.62$ \\
Normalization ($\times10^{-6}$cm$^{-5}$) & $7.81\pm 2.74$ & $11.70\pm 5.13$  & $3.72\pm 0.87$ & $1.52\pm 0.52$ & - & -\\
Emission Measure ($\times10^{-4}$cm$^{-6}$pc) & $4.87\pm 1.71$ & $4.56\pm 1.99$  & - & - & 34 & 37 \\
$\textup{N}_{\footnotesize{\textup{H}}}(\textup{cm}^{-2})$ & $0.0088\times10^{22}$ & $0.0088\times10^{22}$ & $0.0088\times10^{22}$ & $0.0088\times10^{22}$ & $0.159/0.0086\times10^{22}$ & $0.018\times10^{22}$\\
\sidehead{\textbf{Absorbed powerlaw:}}
Normalization\tablenotemark{c} & $6.16 \pm 0.14$ & $5.54 \pm 0.11$  & $1.83 \pm 0.10$ & $7.74 \pm 0.24$ & 14.8 & 12.3\\
Power-law $\Gamma$ & $1.92 \pm 0.0041$ & $1.90 \pm 0.035$ & $2.20 \pm 0.12$ & $1.87 \pm 0.0050$  & 2.2 & 1.52 \\
Reduced $\chi^2$ & 1.06 & 1.06 & 1.02 & 1.21 &  - &  - 
\enddata
%\tablenotetext{a}{This contains the lines O VII and O VIII modeled as $\delta$ functions.}
%\tablenotetext{b}{Galaxy spectra with overlapping areas reduced}
%\tablenotetext{c}{Galaxy spectra without overlapping areas reduced}
\tablenotetext{a}{Galeazzi, M., Gupta, A., Covey, K., \& Ursino, E., et al. 2007, ApJ, 658, 1081. The column densities are toward MBM20 and the Eridanus Hole, respectively.}
\tablenotetext{b}{McCammon, D., Almy, R., Apodaca, E., et al. 2002, Apj, 576, 188}
\tablenotetext{c}{Units are photons s$^{-1}$ keV$^{-1}$ cm$^{-2}$ sr$^{-1}$.}
\end{deluxetable*}
%\end{longrotatetable}
%%%%%%%%%%%%%%%%%%%%%%%%%%%%%%%%%%%%%%%%%%%%--TABLE 3--%%%%%%%%%%%%%%%%%%%%%%%%%%%%%%%%%%%%%%%%%%%%%%%%%%%%%%%%%%
\begin{deluxetable*}{lcccc}[t]
\tablecaption{Emission Measure of the Absorbed Thermal Component}
\label{tab:EM}
\tablehead{
\colhead{ } & \colhead{DXB-90} & \colhead{DXB-50} & \colhead{Galaxy-90} & \colhead{Galaxy-50}} 
\startdata
Solid Angle($10^{-6}$ sr)& 6.55 & 10.48 & 7.61 & 1.59 \\
Model Flux($10^{-6} \times \textup{photons}/\textup{cm}^{2}/\textup{s}$ )& 11.1 & 17.7 & 12.9 & 2.38 \\
Normalization ($\times10^{-6}$cm$^{-5}$) & $7.8\pm 2.8$ & $11.7\pm 5.1$  & $3.7\pm 0.87$ & $1.5\pm 0.52$ \\
Emission Measure ($\times10^{-4}$cm$^{-6}$pc) & $4.9\pm 1.6$ & $4.5\pm 2.0$  & - & - \\
\enddata
\end{deluxetable*}

%%%%%%%%%%%%%%%%%%%%%%%%%%%%%%%%%%%%%%%%%%%%--Ne--%%%%%%%%%%%%%%%%%%%%%%%%%%%%%%%%%%%%%%%%%%%%%%%%%%%%%%%%%%
\begin{deluxetable*}{lcccc}
\tablecaption{Fitting Parameters of Ne IX Component}
\label{tab:NeIX}
\tablehead{
\colhead{Parameter} & \colhead{DXB-90} & \colhead{Galaxy-90} & \colhead{DXB-50} & \colhead{Galaxy-50}}
\startdata
%\sidehead{\textbf{Absorbed plasma component:}}
Line Energy(keV)& $0.93\pm0.008$ & $0.95\pm0.014$ & $0.93\pm0.0084$ & $0.92\pm0.031$ \\
Normalization($\times10^{-6}$ photons cm$^{2}$ s$^{-1}$) & $1.12\pm0.21$ & $0.68\pm0.18$ & $1.33\pm0.26$ & $0.14\pm0.08$ \\ 
Intensity(LU) & $0.17\pm0.032$ & $0.089\pm0.024$ & $0.13\pm0.025$ & $0.088\pm0.05$\\
\enddata
\end{deluxetable*}
%%%%%%%%%%%%%%%%%%%%%%%%%%%%%%%%%%%%%%%%%%%%--Ne--%%%%%%%%%%%%%%%%%%%%%%%%%%%%%%%%%%%%%%%%%%%%%%%%%%%%%%%%%%
%%%%%%%%%%%%%%%%%%%%%%%%%%%%%%%%%%%%%%%%%%%%--AGN--%%%%%%%%%%%%%%%%%%%%%%%%%%%%%%%%%%%%%%%%%%%%%%%%%%%%%%%%%%
\begin{deluxetable*}{lcccccc}
\tablecaption{AGN flux groups parameters}
\label{tab:AGN}
\tablehead{
\colhead{Flux Group\tablenotemark{a}} & \colhead{Temperature(keV)\tablenotemark{b}} & \colhead{Norm\tablenotemark{c}} & \colhead{$\textup{N}_\textup{H}$(cm$^{-2}$)\tablenotemark{d}} & \colhead{$\Gamma$\tablenotemark{e}} & \colhead{Po Norm\tablenotemark{f}} & \colhead{N\tablenotemark{g}}}
\startdata
$10^{-14}$and above & $0.89^{+0.49}_{-0.49}$ & $18.7^{+24.2}_{-24.2}$ & $7.87\times{10^{12}}^{-1.000}_{-1.000}$ & $2.03^{+0.0068}_{-0.0068}$ & $5477.2^{+32.9}_{-32.9}$ & 7 \\
$10^{-15}-10^{-14}$ & $0.16^{+0.020}_{-0.020}$ & $25.9^{+5.63}_{-5.63}$ & $0.046^{+0.0080}_{-0.0080}\times10^{22}$ & $1.66^{+0.016}_{-0.016}$ & $408.9^{+7.61}_{-7.61}$ & 76 \\
$2\times10^{-16}-10^{-15}$ & $0.19^{+0.0041}_{-0.0041}$ & $27.3^{+1.43}_{-1.43}$ & $0.18^{+0.023}_{-0.023}\times10^{22}$ & $1.49^{+0.037}_{-0.037}$ & $100.5^{+4.49}_{-4.49}$ & 134 \\
$10^{-16}-2\times10^{-16}$ & $0.19^{+0.0068}_{-0.0068}$ & $11.8^{+0.85}_{-0.85}$ & $8.72\times{10^{11}}^{-1.000}_{-1.000}$ & $1.25^{+0.036}_{-0.036}$ & $31.2^{+0.92}_{-0.92}$ & 96 \\
$5\times10^{-17}-10^{-16}$ & $0.18^{+0.0056}_{-0.0056}$ & $12.6^{+0.76}_{-0.76}$ & $5.83\times{10^{10}}^{-1.000}_{-1.000}$ & $1.29^{+0.046}_{-0.046}$ & $22.1^{+0.80}_{-0.80}$ & 105 \\
$10^{-17}-5\times10^{-17}$ & $0.19^{+0.0044}_{-0.0044}$ & $13.1^{+0.63}_{-0.63}$ & $1.12\times{10^{12}}^{-1.000}_{-1.000}$ & $1.39^{+0.048}_{-0.048}$ & $17.8^{+0.64}_{-0.64}$ & 149
\enddata
\tablenotetext{a}{Fluxes are in ergs cm$^{-2}$ s$^{-1}$.}
\tablenotetext{b}{Temperature of absorbed thermal component.}
\tablenotetext{c}{Normalization of absorbed thermal component. Units are photons s$^{-1}$ keV$^{-1}$ cm$^{-2}$sr$^{-1}$.}
\tablenotetext{d}{$\textup{N}_\textup{H}$ of absorbed powerlaw.}
\tablenotetext{e}{powerlaw spectral index.}
\tablenotetext{f}{Normalization of absorbed powerlaw. Units are photons s$^{-1}$ keV$^{-1}$ cm$^{-2}$sr$^{-1}$.}
\tablenotetext{g}{Number of AGN in each flux group.}
\end{deluxetable*}
\end{document}